\documentclass[12pt,preprint,eps]{aastex}

\slugcomment{To appear in {\it The Astrophysical Journal}, 558, Sept. 2001}

\shorttitle{Lobel}
\shortauthors{Dynamic Stability of Supergiant Atmospheres}

\begin{document}

\title{On the Dynamic Stability of Cool Supergiant Atmospheres}

\author{A. Lobel}
\affil{Harvard-Smithsonian Center for Astrophysics, 60 Garden Street, Cambridge MA 02138 \\
{\rm (alobel@cfa.harvard.edu)}}

\begin{abstract}
We have developed a new formalism to compute the thermodynamic coefficient $\Gamma_{1}$
in the theory of stellar and atmospheric stability. We generalize the classical derivation 
of the first adiabatic index, which is based on the assumption of thermal ionization and equilibrium 
between gas and radiation temperature, towards an expression which incorporates photo-ionization 
due to radiation with a temperature $T_{\rm rad}$ different from the local kinetic gas temperature.
Our formalism considers the important non-LTE conditions in the extended atmospheres of supergiant stars.   
An application to the Kurucz grid of cool supergiant atmospheres demonstrates that
models with $T_{\rm rad}$$\simeq$$T_{\rm eff}$ between 6500~K and 7500~K 
become most unstable against dynamic perturbations, according to 
Ledoux' stability integral $<$$\Gamma_{1}$$>$. This results from $\Gamma_{1}$ and $<$$\Gamma_{1}$$>$ acquiring very low 
values, below 4/3, throughout the entire stellar atmosphere, which causes very high gas 
compression ratios around these effective temperatures.      
Based on detailed NLTE-calculations, we discuss atmospheric instability of
pulsating massive yellow supergiants, like the hypergiant $\rho$~Cas ($\rm Ia^{+}$), which exist 
in the extension of the Cepheid 
instability strip, near the Eddington luminosity limit.  
\end{abstract}

\keywords{instabilities --- stars: atmospheres --- stellar dynamics --- pulsations --- supergiants 
--- stars: variables: hypergiants --- Cepheids}

\section{Introduction}

The atmospheres of cool massive supergiants are unstable, which causes pulsation-variability,
strongly developed large-scale atmospheric motion fields, excessive mass-loss, and extended 
circumstellar envelopes. One of the best studied examples of these very luminous supergiants 
is the yellow hypergiant $\rho$~Cas (F2$-$G $\rm Ia^{+}$). This evolved star exhibits stable 
pulsation (quasi-) periods of 300$-$500~d. 

Although the $\kappa$- and $\gamma$-mechanisms have been identified as the main cause for 
driving pulsations of the less luminous high-gravity atmospheres of Cepheids, little is known about the 
efficiency of these effects for the much more extended and tenuous atmospheres of cool
massive supergiants. In detailed calculations of the first generalized adiabatic index 
$\Gamma_{1}$, \citet{lob92} (paper I) found that this quantity assumes very small values, below 4/3, 
in low-gravity model atmospheres with 5000~K $\leq$ $T_{\rm eff}$ $\leq$ 8000~K, 
primarily due to the partial thermal ionization of hydrogen. \citet{stc99} recently 
suggested that enhanced mass-loss due to ionization-induced dynamical instability 
of the outer envelope of luminous supergiants which evolve redwards, would terminate 
their redward movement, and provide an explanation for an observational lack of 
yellow and red supergiants with log($L_{\star}$/$L_{\odot}$)$\geq$6.0.  

However, the major problem for evaluating supergiant dynamic (in)stability, based 
on detailed calculations of $\Gamma_{1}$, is the breakdown of LTE conditions in these very extended 
atmospheres. The importance of NLTE ionization- and excitation-conditions is evident from modeling 
the spectra of these stars, which are formed in conditions of very small gravity acceleration. The local ionization equilibrium 
is strongly determined by the stellar radiation field, which determines important
thermodynamic quantities such as the heat capacities, and the related mechanic compressibility of these atmospheres. 

In this paper we develop, for the first time, a self-consistent thermodynamic
formalism which accounts for departures from LTE for the calculation of $\Gamma_{1}$.
This goal is accomplished by introducing the temperature of the radiation field as 
an independent state variable which can differ from the local kinetic gas temperature.
We discuss in Sect. 2 the departure from thermal ionization equilibrium by an incident 
and diluted stellar radiation field in the Eddington approximation. Section 3 
provides a historical overview of the development of the theory of the adiabatic indices. 
The complete analytical expressions for the computation of $\Gamma_{1}$ and the heat capacities in mixtures 
of monatomic gas, interacting with radiation are given in Sect. 4. 
We account for departures from LTE conditions due to the interaction 
of matter and radiation, by evaluating the thermodynamic quantities accordingly.
Section 5 presents a discussion of the effects of NLTE conditions on $\Gamma_{1}$ in cool supergiants.
These detailed NLTE calculations of $\Gamma_{1}$ are applied in Sect. 7, to evaluate 
their dynamic stability according Ledoux' stability integral $<$$\Gamma_{1}$$>$ 
for radial fundamental mode oscillations (Sect. 6). We apply our calculations  
to a new (Kurucz) grid of cool supergiant model atmospheres, which we compute down into 
the stellar envelope. We will demonstrate that NLTE-ionization of hydrogen 
strongly enhances the destabilization of supergiant atmospheres with 6500~K  $\leq$ $T_{\rm eff}$ $\leq$ 7500~K.
For models towards smaller gravity acceleration the stability integral decreases,
and the destabilizing regions occur at lower densities over a larger geometric 
fraction of the atmosphere. A discussion of these results in relation to pulsation 
driving in yellow hypergiants, and atmospheric instability regions recently identified in the 
upper portion of the HR-diagram by \citet{dej97}, is given in Sect.~8. 
The conclusions of this theory and application are listed in Sect.~9.

\section{Non-LTE ionization}

In the atmospheres of supergiants the ionization balance of tenuous gas
is not solely determined by a collisional equilibrium according the Saha-equation.
Significant departures from thermal (equilibrium) ionization occur due to photo-ionization 
by an incident radiation field. 
The local ionization state then becomes dependent of both kinetic gas temperature, and 
the temperature of the radiation field. For example, stellar UV radiation strongly influences the   
Balmer continuum in hot stars.
In deeper atmospheric layers, where the atmosphere is sufficiently optically thick for 
all wavelengths, both temperatures thermalize, and the equilibrium radiation field 
assumes an intensity distribution determined by the local kinetic gas temperature.
In the upper layers the radiation field dilutes with distance from 
a point in the atmosphere where the gas becomes sufficiently optically thin,
and the radiation temperature decouples from the local thermodynamic conditions. 

\subsection{Eddington approximation}
A comprehensive description of partially ionized systems which deviate 
from equilibrium, due to a radiation field of temperature $T_{\rm rad}$, not in equilibrium 
with the electron Maxwell distribution of temperature $T_{\rm e}$, is given in \citet{elw52}.
All particle components (neutrals, ions and electrons) are assumed to be in a Maxwell distribution.  
For the calculation of ionization fractions, this statistical theory assumes 
that the detailed balance between collisional and/or radiative ionization, and recombination processes applies.     
It enables to express the ionization fractions through a departure coefficient $b$ from the Saha equilibrium 
(also the `NLTE Saha-equation'), which can be evaluated using a reduced form of the collision 
ionization cross-section and, for the photo-ionization coefficient, a diluted Planck distribution
with a cross-section obtained from quantum-mechanical calculations.
The Elwert-equation is given by:
\begin{equation}
\frac{n_{j+1}\,n_{\rm e}}{n_{j}} = \left( \frac{n_{j+1}\,n_{\rm e}}{n_{j}} \right)_{\rm Saha} \frac{1}{b_{j}} \,,
\end{equation} 
where \( n_{j} \) denotes the number density of particles in the $j$-th ionization stage,
and $n_{\rm e}$ is the electron number density. The Saha-Boltzmann equation for thermal ionization form the $r$-th 
excitation level is:
\begin{equation}
\left( \frac{n_{j+1}\,n_{\rm e}}{n_{j}} \right)_{\rm Saha} = 2\,\frac{u_{j+1}}{u_{j}} \frac{g_{r,j+1}}{g_{r,j}} 
\left( \frac{2 \pi m_{\rm e}\,k\, T_{\rm e} }{h^{2}} \right)^{\frac{3}{2}} {\rm exp}\left(-\frac{I_{j}-\chi_{r,j}}{k\,T_{\rm e}}\right) \,, 
\end{equation}
with the partition function 
\( u_{j} =\sum_{r=0}^{\infty} g_{r,j}\, {\rm exp}\left( -\frac{\chi_{r,j}}{k\,T_{\rm e}}  \right) \),
and $I_{j}$ the ionization energy from the ground state. \( \chi_{r,j} \) is the excitation energy of 
level $r$ and $g_{r,j}$ its statistical weight. $h$ is the Planck constant  
and the other symbols have their usual meaning. The departure coefficient (for ionization from the ground level) in Eq. (1) is:
\begin{equation}
b_{j} = \frac{ 1+\frac{B}{A\,n_{\rm e}} \frac{1}{y_{\rm e}^{\frac{1}{2}} }\,
F(y_{\rm e},y_{\rm e})}
{1+ W\,\frac{B}{A\,n_{\rm e}} \frac{y^{\frac{1}{2}}_{\rm e}}{y_{\rm rad}}\,
{\rm exp}(y_{\rm e} - y_{\rm rad})\,
F(y_{\rm e},y_{\rm rad})} \,,
\end{equation} 
with the function:
\begin{equation}
F(y_{\rm e},y_{\rm rad}) \simeq  \frac{ \left(  1- \frac{1}{y_{\rm rad}}  \right) } { \left(  1- \frac{2}{y_{\rm e}}  \right) }  \,, 
\end{equation}
where we denote; \( y_{\rm e}= {I_{j}}/{(k\,T_{\rm e})} \) and \( y_{\rm rad}= {I_{j}}/{(k\,T_{\rm rad})} \).
$A$ and $B$ are constants which depend on the ionization energy, the Thompson cross-section, and the Bohr radius.   
Note that Eq. (2) depends only on the kinetic gas temperature $T_{\rm e}$, whereas Eq. (1) 
is dependent of $T_{\rm rad}$ as well.
The dilution factor with geometric height $d$ from the stellar surface $R_{\star}$ is:
\begin{equation}
W(z) = 1/2\,(1 - \sqrt{1 - 1/z^{2}}) \,,
\end{equation}
where $z$=$d$/$R_{\star}$. At the surface $z$=1, the dilution factor $W$($z$=1)=1/2, 
because a gas particle is irradiated at most by half the stellar hemisphere. 

\citet{eck78} distinguished two important conditions of partial ionization from the general 
equation (1), based on a critical electron density $n_{\rm c}$, which is a function of the kinetic temperature,
and the ratio of the radiation and kinetic temperature:
\begin{equation}
n_{\rm c}(T_{\rm e}, T_{\rm rad}/T_{\rm e}) = \frac{B}{A} 
\, \frac{ y_{\rm e}^{\frac{1}{2}} }{ y_{\rm rad} } \, 
{\rm exp}(y_{\rm e} - y_{\rm rad}) \, F(y_{\rm e},y_{\rm rad}) \,.
\end{equation}

$i$. {\it The Corona case} assumes that the radiation density 
is so small that photo-ionization is negligible in comparison 
with electron collision-ionization, and the electron density 
is still so small that three-body recombination is negligible 
in comparison with radiative recombination:
\begin{equation}
n_{\rm c}(T_{\rm e}, 1) \gg n_{\rm e} \gg W\,n_{\rm c}(T_{\rm e}, T_{\rm rad}/T_{\rm e}) \,.
\end{equation}
Hence the departure coefficient can be approximated by:
\begin{equation}
b_{j} \simeq \frac{B}{A\,n_{\rm e}} \frac{1}{y_{\rm e}^{\frac{1}{2}} }\, F(y_{\rm e},y_{\rm e}) = \frac{n_{\rm c}(T_{\rm e},1)}{n_{\rm e}} \,,
\end{equation}
which demonstrates that for coronal conditions the plasma becomes `under'-ionized 
to a degree smaller than the Saha-equilibrium, and for which the ionization fraction 
becomes independent of the local electron number density, since it cancels out in Eq. (1) with Eq. (8).   

$ii$. {\it The Eddington case} assumes that the electron density is so small
that ionization is dominated by photo-ionization, and three-body recombination is dominated
by radiative recombination:
\begin{equation}
n_{\rm c}(T_{\rm e},1) \gg n_{\rm e} \,\,\,\, {\rm and} \,\,\,\, W\,n_{\rm c}(T_{\rm e}, T_{\rm rad}/T_{\rm e}) \gg n_{\rm e} \,.
\end{equation}
Hence, the departure coefficient can be approximated by:
\begin{eqnarray}
b_{j} & & \simeq \frac{1}{W} \frac{y_{\rm rad}}{y_{\rm e}}\,{\rm exp}(y_{\rm rad}-y_{\rm e})\,\frac{F(y_{\rm e}, y_{\rm e})}{F(y_{\rm e}, y_{\rm rad})} \\
      & & = \frac{1}{W} \frac{T_{\rm e}}{T_{\rm rad}} \, {\rm exp}\left( \frac{I_{j}}{k}\left(\frac{1}{T_{\rm rad}} - \frac{1}{T_{\rm e}} \right)\right) \,
\frac{\left(1-\frac{k\,T_{\rm e}}{I_{j}}\right)}{\left(1-\frac{k\,T_{\rm rad}}{I_{j}}\right)}.
\end{eqnarray}
Note that for these conditions of relatively low electron density and high radiation density, the ionization 
fraction still depends on the electron temperature through the radiative recombination mechanism. 

\subsection{Governing ionization equation}

For our calculation of $\Gamma_{1}$ in the atmospheres of supergiants
we consider the Eddington approximation, for which photo-ionization dominates collisional ionization.
The departures from LTE become very large in low density, optically thin, regions.
Because of the radial variation of the local temperature and the wavelength
variation of the opacity sources, the emitted spectrum departs from 
a blackbody at any particular temperature. However, for $T_{\rm eff}$ above 4000~K, 
the local gas density and temperature are only weakly determined by the ambient radiation field, 
because important molecular opacity sources remain limited for these conditions. 
For our calculations we can assume that the stellar radiation field in the atmosphere where 
$\tau_{\rm Ross}$$<$2/3 (i.e. above $R_{\star}$), can be approximated by the effective temperature; $T_{\rm rad}$$\simeq$$T_{\rm eff}$.
For `grey' atmospheres in radiative equilibrium, the color (surface brightness or radiation) temperature is 
0.811 $\times$ $T_{\rm eff}$ \citep{woo53}.

We also consider non-LTE ionization conditions in atmospheric regions where $\tau_{\rm Ross}$$<$2/3,  
with $T_{\rm e}$ and $T_{\rm rad}$ below 20,000~K, for ionization energies $I_{j}$ in excess of 7~eV. 
Hence, the trailing factor at the right-hand 
side of Eq. (11) approaches unity or: 
\begin{equation}
b_{j}  \simeq \frac{1}{W} \frac{T_{\rm e}}{T_{\rm rad}} \, {\rm exp}\left( \frac{I_{j}}{k}\left(\frac{1}{T_{\rm rad}} - \frac{1}{T_{\rm e}} \right)\right) \,. 
\end{equation}
It represents a more tractable expression for our further derivation of important thermodynamic derivatives in Sect. 4.2. 
The NLTE-Saha equation in the Eddington approximation is hence obtained from Eq. (1), Eq. (2), and Eq. (12),
for single ionization ($j$=0$-$1) from the ground level (or \( \chi_{r,j}=0 \)) of element $i$:
\begin{equation}
\frac{n_{i}\,n_{\rm e}}{n_{0}} = 2\, W \, \frac{u_{1,i}}{u_{0,i}} \, 
\left( \frac{2 \pi m_{\rm e}\,k\, T_{\rm rad} }{h^{2}} \right)^{\frac{3}{2}} {\rm exp}\left(-\frac{I_{i}}{k\,T_{\rm rad}}\right)  \,
 \left( \frac{T_{\rm e}}{T_{\rm rad}} \right)^{\frac{1}{2}} \,.
\end{equation}  

\citet{edd26} first considered ionization of the interstellar medium due to ionizing
radiation from nearby stars, and obtained:
\begin{equation}
\frac{n_{i}\,n_{\rm e}}{n_{0}} = 2\, W \, \frac{u_{1,i}}{u_{0,i}} \, 
\left( \frac{2 \pi m_{\rm e}\,k\, T_{\rm rad} }{h^{2}} \right)^{\frac{3}{2}} {\rm exp}\left(-\frac{I_{i}}{k\,T_{\rm rad}}\right) \,,
\end{equation} 
with $T_{\rm rad}$ of the order of the effective temperatures of stellar atmospheres. 
The factor at the right-hand side of Eq. (13), \( \left( T_{\rm e}/T_{\rm rad}  \right)^{\frac{1}{2}}  \) is due to thermal motions, 
and first appears in \citet{ros36}. \citet{str48} gave a further refinement of Eq. (13) 
for photo-ionization from the ground state, which includes recombinations onto energy levels above the 
ground level. \citet{wey62} applied Eq. (13) in a study of the ionization equilibrium 
in the upper atmosphere of the supergiant $\alpha$~Ori (M2 Iab), where $T_{\rm e}$$\leq$5500~K. 
Similar applications to the wind conditions of stellar chromospheres are for example given in \citet{har80}.    

We conclude this section by emphasizing that the condition of detailed balance, for 
photo-ionizations by an equal number of recombinations (per unit volume and unit time) in the 
Eddington approximation, is required for applying Eq. (13).  
Next to the condition of detailed balance, our calculations of 
thermodynamic quantities also require that the kinetic temperatures of the neutrals, 
ions, and electrons thermalize on a time-scale 
shorter than the characteristic time-scale of heat exchange within the fluid; i.e. due to an atmospheric 
temperature gradient. The proton-electron relaxation-time \citep{spi72} is:
\begin{equation}
\ t_{\rm s}({\rm p,e}) = \frac{3\,m_{\rm p}\,(2\pi)^{\frac{1}{2}}\,(k\,T_{\rm e})^{\frac{3}{2}}}
{8\, \pi\, m_{\rm e}^{\frac{1}{2}}\,n_{\rm e}\, Z_{\rm p}^{2}\, e^{4}\, {\rm ln}\,{\Lambda} }
= \frac{503\,A_{\rm p}\, T_{\rm e}^{\frac{3}{2}}}{n_{\rm e}\, Z_{\rm p}^{2}\,\, {\rm ln}\,\Lambda} \,{\rm sec.} \,,
\end{equation}
where \( Z_{\rm p}=1 \)  the proton charge number, \( e \) the electron charge
and \( A_{\rm p} \) the proton mass \( m_{\rm p} \) in atomic units (\( \simeq 1 \)).
For  2$\times$$10^{3}$~K $\leq$\,$T_{\rm e}$\,$\leq$ 2$\times$$10^{4}$~K, in 
the atmospheres of supergiants, where $n_{\rm e}$$\sim$$10^{12}$~$\rm cm^{-3}$, and which are mainly composed of hydrogen, 
we compute for \( {\rm ln}\,\Lambda \simeq 10 \) \citep[as tabulated in][]{spi62}, that large hydrodynamic 
perturbations of the local state variables should not occur on time scales 
shorter than 9$\times$$10^{-7}$~s\,$\leq$\,$t_{\rm s}$\,$\leq$\,5$\times$$10^{-6}$~s, or on characteristic length scales 
smaller than $V_{\rm flow}\,\times\,t_{\rm s}$. 
Equilibrium thermodynamic conditions can for example not establish within the thin  
layer trailing strong shock waves, where the electron temperature departs from the heavy particle gas temperature,
or by the presence of strong electro-magnetic fields which can separate the neutral and charged particle temperatures
of a tenuous plasma. 

\section{Stellar stability coefficients}
Eddington (1918, 1919, 1926) first derived an analytical expression for the first adiabatic index
$\Gamma_{1}$, for a mixture of material gas with radiation. His equation (129.52)
is:
\begin{equation}  
\Gamma_{1} = \beta + \frac{ (4-3\beta)^{2} (\gamma -1)}{\beta + 12(\gamma -1)(1-\beta)} \,,
\end{equation}
where \( \gamma \) is the specific heat ratio, and \( \beta \) the ratio of the material gas 
pressure to the total pressure (using modern symbols). He distinguished
a general, or `effective', ratio of specific heats from the exponent to the density in the polytropic 
equation of state: \( P = K \rho^{\Gamma} \), where $K$ is a constant. 
Strictly thermodynamically speaking, the quantities $\Gamma$ and $\Gamma_{1}$ are not identical, 
but the terminology of `an exponent' to indicate Eddington's adiabatic quantity:
\begin{equation}
\Gamma_{1} \equiv \left( \frac{\partial {\rm\, ln}\, P }{\partial {\rm\, ln}\, \rho } \right)_{\rm ad} 
\end{equation}
has been adopted since in the astrophysical literature. Eddington's consideration of evaluating 
{\it adiabatic} thermodynamic derivatives for studying adiabatic stellar oscillations was 
later more rigorously addressed by \citet{cha39}. He defined the two other  
adiabatic indices:
\begin{equation}
\frac{\Gamma_{2}}{\Gamma_{2}-1}  \equiv \left( \frac{\partial {\rm\, ln}\, P }{\partial {\rm\, ln}\, T } \right)_{\rm ad} \,,
\end{equation}
and
\begin{equation}
\Gamma_{3}-1 \equiv \left( \frac{\partial {\rm\, ln}\, T }{\partial {\rm\, ln}\, \rho } \right)_{\rm ad} \,,
\end{equation}
and also first obtained detailed expressions for a mixture of material gas and radiation. 
$P$ denotes the total pressure, i.e. the sum of partial material gas pressures and radiation pressure. 
Only two of the three 
adiabatic indices are independent because a non-degenerate gas state is determined by at least two 
independent state variables:
\begin{equation}
\Gamma_{3}-1 \equiv \frac{(\Gamma_{2}-1)\,\Gamma_{1}}{\Gamma_{2}} \,.
\end{equation}
With these definitions, Chandrasekhar also obtained more general expressions 
for the specific heats $C_{p}$ and $C_{v}$, and clearly distinguished $\Gamma_{1}$ from
$C_{p}/C_{v}$. His equation (148);
\begin{equation}
\Gamma_{1} = \frac{C_{p}}{C_{v}} \beta \,,
\end{equation} 
where
\begin{equation}
C_{p}= \frac{c_{p}}{\gamma\,\beta^{2}} \left(\beta^{2}+ (\gamma-1)(4-3\beta)^{2}+12(\gamma-1)\beta(1-\beta)   \right) \,,
\end{equation} 
and 
\begin{equation}
C_{v}= \frac{c_{v}}{\beta} \left( \beta + 12 (\gamma -1)(1-\beta)     \right) \,,
\end{equation}
with $c_{p}$ and $c_{v}$ the specific heats of the material gas,
marks an important step towards a self-consistent description of thermodynamic derivatives 
required in the theory of dynamical, convective, and pulsational stability
of stellar atmospheres. \citet{fow25} were the first to derive expressions 
for the specific heats with radiation, where various stages of ionization of the material are allowed.
However, they made certain assumptions about the weight factors and the excitation 
of atoms and ions, which were too restricted. Independently, \citet{mog39}
derived expressions for the specific heats of a singly-ionizing one-component monatomic gas without 
radiation, which were corrected shortly after by \citet{bie42}. He calculated:
\begin{equation}
 c_{p} = \frac{N\,k}{2} (1+x) \left( 5 + x(1-x)\left( \frac{5}{2} + \frac{I}{k\, T}    \right)^{2}     \right) \,,
\end{equation} 
and
\begin{equation}
 c_{v} = \frac{N\,k}{2} (1+x) \left( 3+ \frac{2x}{2-x} \frac{1-x}{1+x}  \left(\frac{3}{2} + \frac{I}{k\,T} \right)^{2}   \right) \,,
\end{equation} 
where $I$ is the ionization energy, $x$ the ionization fraction, and $N$ is the number of atoms per unit mass. 
Biermann also derived the correct expression for the first adiabatic index (evidently, without using this terminology), 
for negligible radiation pressure:
\begin{equation}
\Gamma_{1}= \frac{ 5 + x(1-x)\left( \frac{5}{2} + \frac{I}{k\, T}    \right)^{2}  }{  3 + 
x(1-x) \left( \frac{3}{2} + \left(\frac{3}{2} + \frac{I}{k\,T}     \right)^{2} \right)   } \,.
\end{equation}
Note that the expression simplifies to Eq. (16) without partial ionization ($x$=0), since  $\beta$$\rightarrow$1  
with vanishing radiation pressure, and hence $\Gamma_{1}$$\rightarrow$$\gamma=\frac{5}{3}$ for monatomic gas.
  
\citet{ros48} extended the expressions for the specific heats by considering a mixture 
of partially ionizing hydrogen and helium gas. They also provided a detailed numerical evaluation of their expressions. 
These calculations followed an investigation by \citet{uns38} of Schwarzschild's convection criterion.
He improved upon earlier work by Siedentopf (1933a, 1933b, 1935), and obtained the correct equation for the adiabatic 
temperature derivative: 
\begin{equation}
\frac{\Gamma_{2}}{\Gamma_{2}-1} = \frac{5+x(1-x)\left( \frac{5}{2} + 
\frac{I}{k\,T} \right)^{2}}{2+x(1-x)\left( \frac{5}{2} + \frac{I}{k\,T} \right)} \,.
\end{equation}  
\citet{uns38} introduced the `mean' degree of ionization \( \bar{x}=\sum_{i}^{m} \nu_{i} x_{i} \),
with $\nu_{i}$ the abundance of element $i$, and hence $\bar{x}=1$ for a fully ionized gas. 
It enabled to obtain an extended analytical 
expression for  \( \frac{\Gamma_{2}}{\Gamma_{2}-1} \) in terms of summations over the ionization 
fractions of $m$ elements of a gas mixture [see his equation (93,20)]. 
In the second edition of his monograph on stellar atmospheres, \citet{uns68} also derived an expression for $c_{p}$,
by means of the definition of $\bar{x}$. A similar treatment for $c_{v}$ is given in \citet{men63}.  
These equations however omit the important influence of a radiation field, as Chandrasekhar demonstrated. 
This problem has been investigated by \citet{kri63}, who computed the adiabatic temperature derivative 
for a multi-component mixture of singly-ionizing monatomic gas and 
radiation, using an equation of state of the form: \( P_{\rm t} = N\,k\,T\,(1+\bar{x})\rho + \frac{1}{3}\,a\,T^{4} \). 
The complete analytical expressions for the specific heats of this mixture were first derived 
by \citet{mih65}:
\begin{eqnarray}
\frac{c_{P_{\rm t}}}{N\,k} & & = \left( \frac{5}{2} + 20\alpha + 16 \alpha^{2} \right)(1+\bar{x})
+ \sum_{i}\nu_{i}x_{i}(1-x_{i}) \left( \frac{I_{i}}{k\,T}  \right)^{2}  \nonumber \\
 & & + \left(\frac{5}{2} + 4\alpha \right) \sum_{i}\nu_{i}x_{i}(1-x_{i})\frac{I_{i}}{k\,T}  
+ \left( \frac{5}{2} + 4\alpha + \frac{\sum_{i} \nu_{i}x_{i}(1-x_{i})\frac{I_{i}}{k\,T})}{\bar{x}- < x^{2}>} \right)  \nonumber \\
& & \times \frac{\bar{x} - <x^{2}> }{ \bar{x}^{2} - 2\bar{x} - <x^{2}>}
 \left( \bar{x} (1+\bar{x})\left(\frac{5}{2}+4\alpha\right) - \sum_{i} \nu_{i} x_{i} 
(1 - x_{i}) \frac{I_{i}}{k\,T} \right) \,,
\end{eqnarray}
and 
\begin{eqnarray}
\frac{c_{v}}{N\,k} & & = \left( \frac{3}{2} + 12\alpha \right)(1+\bar{x})
+\frac{3}{2} \sum_{i}\nu_{i}x_{i}(1-x_{i}) \left( \frac{I_{i}}{k\,T}  \right) 
+ \sum_{i}\nu_{i}x_{i}(1-x_{i}) \left( \frac{I_{i}}{k\,T}  \right)^{2}  \nonumber \\
& & + \left( \frac{3}{2}\bar{x} - \sum_{i} \nu_{i}x_{i}(1-x_{i})\frac{I_{i}}{k\,T} \right)
\frac{\frac{3}{2}(\bar{x}-<x^{2}>) + \sum_{i} \nu_{i} x_{i} (1-x_{i})\frac{I_{i}}{k\,T}}{ 2\bar{x} -<x^{2}> } \,,
\end{eqnarray}
where \( <x^{2}>=\sum_{i} \nu_{i} x_{i}^{2} \), $I_{i}$ is the ionization energy of element $i$,
$\alpha$ is the ratio of the radiation pressure and the material gas pressure, and the other symbols have their usual meaning.
These equations simplify to Eq. (24) and Eq. (25) in case the radiation pressure vanishes
($\alpha \rightarrow 0$), for a one-component gas $m$=1 (hence $\bar{x}=x$ and $<x^{2}>=x^{2}$). 
They also simplify to 
Eq. (22) and Eq. (23) when partial ionization of all elements vanishes ($x_{i} \rightarrow 0$), since 
$\alpha = 1/\beta -1$, and for monatomic ideal gas $c_{p} = 5/2\, N\,k$, $c_{v}=3/2\,N\,k$, or $\gamma=c_{p}/c_{v}=5/3$.
The analytical representation of Eq. (28) and Eq. (29) appears rather complex, but we
will show in Sect. 4.2 that their further generalization, in which the gas temperature differs from 
the radiation temperature, enables to define for every element $i$ two functions $G_{i}$ and $H_{i}$,   
which reduce $c_{P_{\rm t}}$ and $c_{v}$ to basically two terms. 

The calculation of the adiabatic indices for real gas in the context of equilibrium thermodynamics  
was first given by \citet{cox68}. 
The specific heat ratio is related to the ratio of the isothermal and adiabatic compressibility coefficient 
via the Maxwell relations:
\begin{equation}
\frac{c_{p}}{c_{v}} = \frac{\kappa_{T}}{\kappa_{S}},
\end{equation}  
with the coefficients:
\begin{equation}
\kappa_{T} \equiv \left( \frac{ \partial {\rm\, ln}\, \rho }{ \partial\, P   } \right)_{T} \,\,\, {\rm and} \,\,\, 
\kappa_{S} \equiv \left( \frac{ \partial {\rm\, ln}\, \rho }{ \partial\, P   } \right)_{S} \,,
\end{equation}
where $S$ denotes the thermodynamic entropy function.
Hence, $\Gamma_{1}$ can be expressed by:
\begin{equation}
\Gamma_{1} \equiv \left( \frac{ \partial {\rm\, ln}\, P_{\rm t}  }{ \partial {\rm\, ln}\, \rho} \right)_{S} 
= \frac{c_{p_{\rm t}}}{c_{v}} \left( \frac{\partial {\rm\, ln}\, P_{\rm t}  }{ \partial {\rm\, ln}\, \rho } \right)_{T} \,.
\end{equation}
The thermodynamic derivative \( \left(  \frac{\partial\, {\rm ln}\, P_{\rm t}  }{ \partial\, {\rm ln}\, \rho }  \right)_{T} \),
is the $\beta$-factor at the right-hand-side of Eq. (21), which Chandrasekhar distinguished from the specific heat ratio
by calculating Eddington's index for a mixture of ideal material gas and radiation. 
\citet{cox68} proposed to term this factor \( \left(  \frac{\partial\, {\rm ln}\, P_{\rm t}  }{ \partial\, {\rm ln}\, \rho }  \right)_{T} \)
the ``density exponent in the pressure equation of state'' $\chi_{\rho}$,
probably following the adopted designation of `adiabatic exponent' for $\Gamma_{1}$;
\begin{equation}
\chi_{\rho} \equiv \left(  \frac{\partial\, {\rm ln}\, P_{\rm t}  }{ \partial\, {\rm ln}\, \rho }  \right)_{T} \,,\,\,\, {\rm hence}\,\,\, 
\Gamma_{1}=\frac{c_{p_{\rm t}}}{c_{v}}\, \chi_{\rho} \,.
\end{equation}
Although this terminology has widely been adopted in the literature, we note
that this quantity is by no means a true `exponent' in the polytropic equation of state. It is merely 
a coefficient to the specific heat ratio, required to determine an important adiabatic derivative. 
\citet{fow25} also calculated this quantity and aptly called it `the isothermal factor' (see their Eq. 9), 
since it is inversely proportional to the isothermal compressibility coefficient: 
\( \kappa_{T} = (P_{\rm t}\, \chi_{\rho})^{-1} \).  
\citet{cox68} also introduced the ``temperature exponent'' (more adequately `the isochoric factor'):
\begin{equation}
\chi_{T} \equiv \left(  \frac{\partial\, {\rm ln}\, P_{\rm t}  }{ \partial\, {\rm ln}\, T }  \right)_{\rho} \,,
\end{equation}      
which provides an important general relation for the thermodynamics of real gas:
\begin{equation}
c_{p_{\rm t}} - c_{v} = \frac{P_{\rm t}}{\rho \, T} \, \frac{\chi^{2}_{T}}{\chi_{\rho} } \,.
\end{equation}
Thermodynamic stability demands that both heat capacities and both compressibilities
are positive. For every real gas \( c_{P_{\rm t}} > c_{v}\) and \( \kappa_{T} > \kappa_{S} > 0    \),
and \( c_{P_{\rm t}} - c_{v} \)  does not equal the universal gas constant.  The latter
equality applies only to simple ideal gas, for which \( \chi_{T} = \chi_{\rho} = 1 \).

With the definitions of the isothermal and isochoric factor, the two other adiabatic indices 
are obtained with:
\begin{equation}
\frac{\Gamma_{2}}{\Gamma_{2}-1} = \frac{c_{P_{\rm t}}}{c_{P_{\rm t}}-c_{v}} \, \chi_{T} \,,
\end{equation} 
and 
\begin{equation}
\Gamma_{3} - 1 = \frac{c_{P_{\rm t}}-c_{v}}{c_{v}} \, \frac{\chi_{\rho}}{\chi_{T}} \,,
\end{equation}
which yields the general identity Eq. (20). The complete expressions for $\chi_{\rho}$ 
and $\chi_{T}$ for a singly-ionizing multi-component monatomic gas with radiation 
are \citep{lob92}:
\begin{equation}
\chi_{\rho} = \frac{\beta\,\left( \bar{x}^{2}+\bar{x}+\sum_{i} \nu_{i} x_{i} (1-x_{i})\right)}
{(1+\bar{x})\left( \bar{x} + \sum_{i} \nu_{i} x_{i} (1-x_{i})    \right)   } \,,
\end{equation}
and
\begin{equation}
\chi_{T}= (4-3\,\beta) + \frac{\beta\,\bar{x} \sum_{i} \nu_{i} x_{i} (1-x_{i}) (\frac{3}{2}+\frac{I_{i}}{k\,T})}
{(1+\bar{x})\left(\bar{x}+\sum_{i} \nu_{i} x_{i} (1-x_{i})      \right)    }  \,.
\end{equation}
Note that for a hypothetically non-ionizing gas ($x_{i}$,~$\bar{x}$)$\rightarrow$0, whence $\chi_{\rho}$$\rightarrow\beta$ and
$\chi_{T}$$\rightarrow$4$-$3$\beta$.
These expressions simplify to those of \citet{cox68} for a gas composed of one ionizing element ($\bar{x}=x$) 
and radiation. Note that \citet{bie42} first derived $\chi_{\rho}$ without radiation ($\beta=1$), however expressed as a factor
to \(c_{P_{\rm t}}/c_{v}  \):
\begin{equation}
\chi_{\rho} \equiv
\left( \frac{\partial\, {\rm ln}\, P_{\rm t}}{ \partial\, {\rm ln}\, \rho} \right)_{T} = 1 - 
\left( \frac{\partial\, {\rm ln}\, \mu}{ \partial\, {\rm ln}\, \rho} \right)_{T} = \frac{2}{(1+x)(2-x)} \,,
\end{equation}
where $\mu$ is the mean molecular weight. In the partial ionization zone of an abundant element 
the mean molecular weight reduces because \(\mu = \mu_{0}/(1+\bar{x}) \), where $\mu_{0}$ is the 
mean molecular weight of the unionized gas (e.g. $\mu_{0}$=1.26 for the abundance of cosmic material).
In these regions the mean ionization fraction $\bar{x}$ increases ($0 \leq \bar{x} \leq 1$) and, 
as can be seen from Eq. (38), $\chi_{\rho}$ assumes values below unity. The increased compressibility in the thermal ionization 
regions reduces with Eq. (32) the value of the first adiabatic index $\Gamma_{1}$ to below the monatomic 
gas value of 5/3. Here, compression energy is mainly converted into ionization energy, which also changes the 
local heat capacities. An equilibrium radiation field reduces $\beta\rightarrow 0$, and hence $\chi_{\rho}\rightarrow 0$ 
and $\chi_{T} \rightarrow 4$. 

\citet{lob92} demonstrated that $\Gamma_{1}$ can assume values {\it below unity}  
when isotropic radiation pressure is important in the partial ionization region of an abundant element, although 
$\gamma$=$c_{P_{\rm t}}/c_{v}>1$ for every stable gas. The higher compressibility of radiation, compared to pure material gas 
of the same temperature and pressure, diminishes $\chi_{\rho}$ to very low values for 6100~K$\leq$ $T_{\rm e}$ $\leq$9000~K, in the 
partial ionization region of hydrogen. Hence, $\Gamma_{1}$ can decrease to very small values of 0.84. 
Note however that when radiation vanishes,
the product of the specific heat ratio and $\chi_{\rho}$ always exceeds unity ($\Gamma_{1}$$\geq$1).
The extended Eqns. (28), (29), (38), and (39) also enable to correctly compute the compressibility of atmospheric regions
where simultaneous ionizations of various elements (e.g. $\rm H\rightarrow H^{+}$ and $\rm He \rightarrow He^{+}$) 
considerably lower the values of $\Gamma_{1,\,2,\,3}$. 

The combination of ionization and radiation can diminish $\Gamma_{1}$ to below 4/3, which plays an important 
role in the study of dynamic stability of gas spheres, which we discuss in Sect. 6. 
Conventional calculations of the adiabatic indices assume however that the gas and radiation temperature are equal, and
that no interaction occurs between the material gas and the radiation field. 
We presently investigate the effects of a radiation field on the adiabatic indices 
due to photo-ionization in the Eddington approximation. The local ionization state 
is no longer determined by pure collisional processes, but by the incident and diluted radiation field as well. 
Radiative deviations from the Saha equilibrium produce important effects on the overall compressibility of the plasma,
which directly determines the dynamic stability of supergiant atmospheres.
             
\section{First adiabatic index with photo-ionization}
\subsection{Definition of the generalized functions}
When the radiation temperature differs from the local kinetic gas temperature, the classic equation
of state, which assumes thermal equilibrium between material gas and radiation 
\( P_{\rm t} = N\,k\,T\,(1+\bar{x})\rho + \frac{1}{3}\,a\,T^{4} \), is replaced by:
\begin{equation}
P_{\rm t} =  N\,k\,T_{\rm e}\,(1+\bar{x})\rho + \frac{W}{3}\,a\,T_{\rm rad}^{4} \,,
\end{equation}
where the first term is the material gas pressure \( P_{\rm g}   \),
and the second term the diluted radiation pressure \( P_{\rm rad} \).
The gas state is hence determined by three independent state variables, instead of two.
Therefore, the calculation of the first adiabatic index must consider $T_{\rm e}$ and $T_{\rm rad}$ 
as independent state variables. The constituent heat capacities           
\begin{equation}
c_{P_{\rm t}}= \left( \frac{\partial\, h}{\partial\, T} \right)_{P_{\rm t}}, \,\,\, {\rm and} \,\,\,
c_{v}    = \left( \frac{\partial\, e}{\partial\, T} \right)_{\rho},
\end{equation}  
are replaced by:
\begin{equation}
c_{P_{\rm t}}= \left( \frac{\partial\, h}{\partial\, T_{\rm e}} \right)_{P_{\rm t},\, T_{\rm rad}} 
+ \left( \frac{\partial\, h}{\partial\, T_{\rm rad}} \right)_{P_{\rm t},\, T_{\rm e}} \,,
\end{equation}
and
\begin{equation}
c_{v}= \left( \frac{\partial\, e}{\partial\, T_{\rm e}} \right)_{\rho,\, T_{\rm rad}} 
+ \left( \frac{\partial\, e}{\partial\, T_{\rm rad}} \right)_{\rho,\, T_{\rm e}} \,,
\end{equation}
where $e$ denotes the internal energy function, and $h$ is the enthalpy function. The lower case symbols
denote `specific' quantities, or expressed per unit mass. 
The isothermal factor
\begin{equation}
\chi_{\rho} = \left( \frac{\partial\, {\rm ln}\, P_{\rm t}}{\partial\, {\rm ln}\, \rho} \right)_{T} \,,
\end{equation}
is replaced by:
\begin{equation}
\chi_{\rho} = \frac{1}{P_{\rm t}} \left( \left( \frac{\partial\, P_{\rm g}}{ \partial\, {\rm ln}\, \rho }  \right)_{T_{\rm e}}
+ \left( \frac{\partial\, P_{\rm rad}}{ \partial\, {\rm ln}\, \rho }  \right)_{T_{\rm rad}    }     \right) \,.
\end{equation}
It is important to note that the three adiabatic indices become fully independent by the introduction of an additional 
state variable $T_{\rm rad}$. Since the second and third adiabatic indices are temperature derivatives they are 
also re-defined by:
\begin{equation}
\frac{\Gamma_{2\,\rm g}}{\Gamma_{2\,\rm g}-1} \equiv \left(\frac{\partial\, {\rm ln}\, P_{\rm t} }{ \partial\, {\rm ln}\, T_{\rm e} } \right)_{\rm ad} \,, \,\,\, {\rm and} 
\,\,\, \frac{\Gamma_{2\,\rm rad}}{\Gamma_{2\,\rm rad}-1}  \equiv \left(\frac{\partial\, {\rm ln}\, P_{\rm t} }{ \partial\, {\rm ln}\, T_{\rm rad} } \right)_{\rm ad} \,,
\end{equation} 
and by
\begin{equation}
\Gamma_{3\,\rm g} -1 \equiv \left(\frac{\partial\, {\rm ln}\, T_{\rm e} }{ \partial\, {\rm ln}\, \rho } \right)_{\rm ad} \,, \,\,\, {\rm and} 
\,\,\, \Gamma_{3\,\rm rad} -1 \equiv \left(\frac{\partial\, {\rm ln}\, T_{\rm rad} }{ \partial\, {\rm ln}\, \rho } \right)_{\rm ad} \,.
\end{equation}
With these definitions the identity Eq. (20) no longer applies, because it is only valid for thermodynamic 
systems with a unique temperature. 

\subsection{Derivation of the generalized functions}
In the Appendix we obtain the detailed expression for the specific heat capacities for which $T_{\rm e}$  
differs from $T_{\rm rad}$. The appendix is self-contained and can be read without further cross-references. 
All thermodynamic quantities and functions are defined, 
and their detailed expressions are presented there. Major intermediate results, required to obtain the detailed 
expression for $c_{P_{\rm t}}$ and $c_{v}$, are also provided. 
We denote $\theta = T_{\rm e}/T_{\rm rad}$, and find after considerable algebra:
\begin{equation}
\frac{c_{P_{\rm t}}}{Nk} = \left( \frac{5}{2} + 4\,\alpha\,\left( 4\,\theta\,(\alpha+1)+1 \right) \right)  (1+\bar{x})
+ \sum_{i} \nu_{i} x_{i} (1-x_{i})\, H_{i} \,,
\end{equation}
and 
\begin{equation}
\frac{c_{v}}{N\,k} = \left( \frac{3}{2} + 
12\,\alpha\,\theta \right)(1+\bar{x}) + \sum_{i} \nu_{i} x_{i} (1-x_{i}) \,G_{i} \,, 
\end{equation}  
with the functions:
\begin{eqnarray} 
X=& & \sum_{i} \nu_{i} x_{i} (1-x_{i}) \,, \\
Y = & & \sum_{i} \nu_{i} x_{i} (1-x_{i}) \frac{I_{i}}{k\, T_{\rm e}} \,,
\end{eqnarray}
and for every element $i$: 
\begin{eqnarray}
Q_{i} = & & \frac{1}{2} +  \theta \left(1+ \frac{I_{i}}{k\, T_{\rm rad}} \right) \,, \\
G_{i} = & & Q_{i} \left( \frac{\frac{3}{2}\bar{x}-Y}{\bar{x}+X} + \frac{I_{i}}{k\,T_{\rm e}}  \right) \,, \\
H_{i} = & &  \left( Q_{i} + 1 + 4\,\alpha\,\theta \right) \left( \frac{ (\frac{5}{2} + 4\,\alpha)\, \bar{x}\, (1+\bar{x}) - Y}
{ \bar{x}\, (1+\bar{x}) + X} + \frac{I_{i}}{k\,T_{\rm e}}\right) \,.
\end{eqnarray}
It can be shown that for $\theta=1$, Eq. (49) and Eq. (50) simplify to Eq. (28) and Eq. (29).
The first term in Eq. (49) and Eq. (50) is the heat capacity due to the translational motion of neutral atoms, ions, 
and electrons, and $\alpha$ is dependent of the diluted radiation pressure. 
The second term increases the heat capacities due to extra internal degrees of freedom, which results 
from partial photo-ionization by the incident radiation field. When 
$T_{\rm rad}\rightarrow T_{\rm e}$, the NLTE-ionization balance assumes the thermal Saha-equilibrium, and 
the generalized heat capacities simplify to the heat capacities for a unique temperature $T$.    

The analytical expression for the isothermal factor $\chi_{\rho}$, in which $T_{\rm rad}$ differs from 
$T_{\rm e}$, is formally identical with Eq. (38):
\begin{equation}
\chi_{\rho} = \frac{\beta\,\left( \bar{x}^{2}+\bar{x}+\sum_{i} \nu_{i} x_{i} (1-x_{i})\right)}
{(1+\bar{x})\left( \bar{x} + \sum_{i} \nu_{i} x_{i} (1-x_{i})    \right)   } \,.
\end{equation}
However, the dilution of the radiation pressure enters this generalized 
expression through $\beta$.
The second term at the right-hand-side of Eq. (46) vanishes because $P_{\rm rad}$ is invariable 
for constant $T_{\rm rad}$, and $P_{\rm rad}$ is independent of $T_{\rm e}$. The 
derivative in the first term is evaluated for variable $T_{\rm rad}$, because the radiation temperature determines 
the NLTE-ionization fraction.
    
In Sect. 7 we  
investigate numerically the properties of the NLTE $\Gamma_{1}$ for multi-component monatomic gas with 
radiation. If we consider only one element of material gas ($m$=1 and $\bar{x}=x$), Eq. (49) simplifies to \citep[see][]{lob97}:
\begin{eqnarray}
\frac{c_{P_{\rm t}}}{N\,k} = & &  (1+x)\,\left\{ \left( \frac{5}{2}+ 4\,\alpha\,(4\,\theta\,(\alpha+1)+1) \right) \right. \nonumber \\
+ & &  \frac{x\,(1\,-\,x)}{2} \left( \frac{5}{2}\!+\!\frac{I}{k\,T_{\rm e}}\!+\!4\, \alpha \right)
\left. \left( \frac{3}{2}\!+\!\theta\left(1\!+\!\frac{I}{k\,T_{\rm
rad}}\right)\!+\!4 \, \alpha\,\theta \right) \right\} \,,
\end{eqnarray}
and Eq. (50) to:
\begin{equation}
\frac{c_{v}}{N\,k} =  \left( \frac{3}{2}+12\,\alpha\,\theta \right) \,(1+x) 
+ \frac{x(1-x)}{2-x}
\left( \frac{3}{2}+\frac{I}{k\,T_{\rm e}} \right)
\left(\frac{1}{2}+\theta\left(1+\frac{I}{k\,T_{\rm rad}}\right) \right) \,.
\end{equation}
Because the isothermal factor Eq. (56) simplifies to:
\begin{equation}
\chi_{\rho} = \frac{2\,\beta}{(1+x)(2-x)} \,,
\end{equation}
we obtain after some algebra the NLTE `one-component' first adiabatic index:
\begin{eqnarray}
\Gamma_{1} = & & \frac{c_{P_{\rm t}}}{c_{v}} \chi_{\rho}  \nonumber \\
= & & \beta \, \frac{\frac{5}{2}+4\alpha(4\theta(\alpha+1)+1) +
\frac{x}{2}(1-x)\left(\frac{3}{2} +
\Phi + 4\alpha\right)\left(\frac{3}{2}+
\theta\Psi+4\alpha\theta\right)}{\frac{3}{2}+
12\alpha\theta+\frac{x}{2}(1-x)\left(\left(\frac{1}{2}+ \Phi
\right)\left(\frac{1}{2}+
\theta\Psi \right) +\frac{3}{2}+12\alpha\theta \right)} \,,
\end{eqnarray}
where we denote:
\begin{equation}
\Phi = 1 + \frac{I}{k\,T_{\rm e}} \,\,\,  {\rm and} \,\,\,  \Psi = 1 + \frac{I}{k\,T_{\rm rad}} \,.
\end{equation}
For $T_{\rm e}$=$T_{\rm rad}$=$T$, hence $\theta=1$ and \( \Phi = \Psi =1+\frac{I}{k\,T} \), it follows that 
Eq. (60) simplifies to \citep[e.g. Eq. (71.16) of][]{mih84}:
\begin{equation}
\Gamma_{1} = \beta\,\frac{\frac{5}{2} + 20\alpha + 16 \alpha^{2} + \frac{x}{2}(1-x) \left( \frac{5}{2} + \frac{I}{k\,T} + 4\,\alpha  \right )^{2}}
{\frac{3}{2} + 12 \alpha + \frac{x}{2}(1-x) \left( \left(\frac{3}{2} + \frac{I}{k\,T}\right)^{2} + \frac{3}{2} + 12\alpha   \right)    } \,,
\end{equation}
which simplifies without radiation ($\alpha$=0 and $\beta$=1) to Eq. (26).    

We have also obtained the detailed expressions for the two other adiabatic indices $\Gamma_{2}$ and $\Gamma_{3}$
along similar lines, defined with Eq. (47) and Eq. (48), but which will be presented elsewhere. 
These generalized functions do not simplify to the classic expressions derived 
for a unique $T$, because the partial {\em temperature} derivatives, computed with  
the Saha equilibrium equation, differ from those obtained using the NLTE-Saha equation. It should be 
remembered that the latter is an approximate expression for which the condition of detailed 
balance is assumed to sustain the condition of (ionization-) equilibrium. The (LTE) Saha-equation, however, is exact 
and follows from the minimization of the free energy function, which dictates the condition of thermodynamic 
equilibrium. 

Note that for the computation of thermodynamic quantities
free energy minimization techniques have been applied to include the effects of
pressure ionization due to Coulomb forces, and the effect of degenerate states
\citep[see][]{hum88, mih88, dap88}. 
This numerical approach offers the advantage, for example, to incorporate the temperature dependency 
of the (properly truncated) internal partition functions, and to compute their effect on the heat 
capacities and the compressibilities. 
However, the free energy minimization scheme is strictly numerical because a large system of non-linear
equations is to be solved iteratively for astrophysical mixtures \citep[e.g.][]{mih90}.

In contrast, our formal approach for mixtures with simultaneously ionizing elements and $T_{\rm rad}$$\neq$$T_{\rm e}$,
offers the advantage to track analytically the complex behavior of the adiabatic indices.
For example, the remarkable occurrence of conditions where $\Gamma_{1}$$<$1. 
Our formalism applies however to supergiant atmospheres, for which electrostatic interactions 
and degenerate states are negligible. Possibly, relativistic or degenerate gas effects on $\Gamma_{1}$ can be incorporated by 
the analytical approximations outlined in \citet{cox68}, \citet{bea71}, 
\citet{ell98}, and \citet{sto00}. \citet{eck63} presented an interesting 
statistical method, based on the minimum entropy production principle of irreversible thermodynamics, to compute 
the lowering of ionization energy for a Saha-type equation which considers Coulomb interaction.
Note also that \citet{mol89} computed effects of a uniform magnetic field 
on the adiabatic indices, and provided generalized LTE expressions for the heat capacities, 
including the magnetic pressure. 

\subsection{General thermodynamic relation for $\Gamma_{1}$}
We conclude this section with an important thermodynamic relation for $\Gamma_{1}$.
From the first law the relationship among $P_{\rm t}$, $\rho$, and $e$ along an isentrope ($d$$S$=0) is:
\begin{equation}
\left( \frac{d\, {\rm ln}\, e }{d\, {\rm ln}\, \rho} \right)_{\rm ad} = \frac{h}{e} -1 \,,
\end{equation}  
where \( h = e + P_{\rm t}/\rho \).
The introduction of an additional state variable $T_{\rm rad}$ defines surfaces of constant (total) entropy with the specific 
entropy function $s$=$s$($\rho$, $T_{\rm e}$, $T_{\rm rad}$) in the space of three independent state variables.
For adiabatic changes:  $d$$e$= $P_{\rm t}$/$\rho^{2}$ $d$$\rho$  and $d$$h$ = 1/$\rho$ $d$$P_{\rm t}$,
the first adiabatic index can be expressed by:
\begin{equation}
\Gamma_{1} \equiv \left( \frac{d\, {\rm ln}\, P_{\rm t}}{ d\, {\rm ln}\, \rho }\right)_{\rm ad}   
= \left( \frac{d\, h}{d\, e} \right)_{\rm ad} \,.
\end{equation}
The value of $\Gamma_{1}$ is related with the slope along an adiabate in the ($P_{\rm t}$, $\rho$)-plane,
for a given $T_{\rm rad}$ or $T_{\rm e}$. Since the NLTE-Saha equation considers 
local interactions of material gas with the incident radiation field, the total entropy function
cannot classically be obtained by summing over the particle entropy fractions, 
the entropy of released electrons, and the entropy of equilibrium radiation 
\citep[i.e. by means of the Sackur-Tetrode equation; see][]{lob97}. 
$s$($\rho$, $T_{\rm e}$, $T_{\rm rad}$) is to be obtained 
from the temperature derivative of the free energy, defined for the gas state  
determined by $T_{\rm e}$ {\it and} $T_{\rm rad}$. 
Similarly, for conditions of equilibrium, $e$=$e$($\rho$, $T_{\rm e}$, $T_{\rm rad}$) 
and $h$=$h$($\rho$, $T_{\rm e}$, $T_{\rm rad}$) are
uniquely defined functions, and it can be shown that their ratio \( \gamma_{\rm H}= h/e \)
is related to the first adiabatic index by:
\begin{equation}
\Gamma_{1} = \gamma_{\rm H} + \frac{\gamma_{\rm H}}{\gamma_{\rm H}-1} 
\left( \frac{d\,{\rm ln\, \gamma_{\rm H}}}{d\,{\rm ln}\, \rho} \right)_{\rm ad} \,.
\end{equation}
Hence, the decrease of $\Gamma_{1}$ due to partial ionization and the release of bound electrons 
is related with the decrease of $\gamma_{H}$ \citep[e.g.][]{nie93}.
While the value of $\Gamma_{1}$ is related with the adiabatic compressibility of the fluid 
(e.g. the bulk modulus) by local mechanic perturbations, the value of $\gamma_{\rm H}$ is 
determined by the corresponding changes of the internal energy. 

\section{Behavior of $\Gamma_{1}$ in supergiant atmospheres}
Figure~1 shows the behavior of $\Gamma_{1}$ and the mean ionization fraction $\bar{x}$,
for variable $T_{\rm e}$ and $T_{\rm rad}$.
We compute the thermodynamic quantities for a mixture of 16 elements comprising
H, He, C, N, O, Ne, Na, Mg, Al, Si, S, Ar, K, Ca, Cr, and Fe, for solar abundance 
values \citep{and89}. The internal partition functions 
are derived according the methods developed by \citet{cla51} and by 
\citet{bas66}. The left-hand panels display the NLTE $\Gamma_{1}$ computed 
at the stellar radius ($W$=1/2), and the right-hand panels show $\bar{x}$. 
In the upper panels the material gas pressure is set to 10~dyn $\rm cm^{-2}$, whereas
0.5~dyn $\rm cm^{-2}$ is used for the lower panels. These conditions are typical
for the atmospheres of cool supergiants. Figure 1 reveals that 
$\Gamma_{1}$ acquires minimum values for $T_{\rm rad}$ between 7000~K and 9000~K,
primarily due to the partial ionization of hydrogen. Towards very low $T_{\rm rad}$,
$\Gamma_{1}$ approaches the value of 5/3 for monatomic ideal gas, whereas for 
high $T_{\rm rad}$ $\Gamma_{1}$ assumes the equilibrium radiation value of 4/3.  
The intermediate regions with lowest $\Gamma_{1}$ correspond to $\bar{x}$$\simeq$0.5. 
The NLTE-ionization equation causes a strong dependence of the local mean ionization 
fraction on the value of $T_{\rm rad}$. For a given radiation temperature, however at 
constant gas pressure, the ionization fraction {\it decreases} towards higher kinetic
gas temperatures, because the NLTE-ionization fraction remains dependent 
of the local kinetic temperature through the factor $T^{\frac{1}{2}}_{\rm e}$ in Eq. (13)
(but which is $T^{\frac{3}{2}}_{\rm e}$ for pure collisional ionization).  

We find minimum values for $\Gamma_{1}$
which can decrease to below 0.7 for very small gas pressures and low $T_{\rm e}$.
The minima gradually increase for larger gas pressures, by 
shifting towards higher $T_{\rm rad}$. 
This is also shown by the $\Gamma_{1}$-surface plots of Fig.~2. 
The deep minimum around $T_{\rm rad}$$\simeq$8000~K ({\em lower panel}) becomes 
shallower and broadens for increased kinetic pressures ({\em upper panel}).    
The relative decrease of the radiation pressure also enhances the 
partial ionization of He, which is visible for $T_{\rm rad}$ between 
14,000~K and 16,000~K. Note that we have also included the partial 
ionization of $\rm He^{+}$ in our calculations. Its ionization 
can be treated as a separate `element', because the ionization energy 
is very high, and partial ionization occurs around 25,000~K 
for $P_{\rm g}$=1~dyn~$\rm cm^{-2}$. Electrons from less abundant metal atoms,
with smaller ionization energies, contribute only slightly 
to the total electron pressure. The much larger abundance of hydrogen 
(by about a factor of ten larger than He) causes the large decrease of $\Gamma_{1}$
to occur for $T_{\rm rad}$ between 7000~K and 9000~K, for the small gas pressure 
conditions of supergiant atmospheres.

An important influence on the local mean ionization fraction 
is the dilution of radiation pressure with distance above $R_{\star}$, which we further discuss in Sect. 7.1.    
Calculations with $W(z)$$<$1/2 (for $z>1$) reveal 
that the minima of $\Gamma_{1}$ increase, which is expected when kinetic 
pressure dominates the gas state.      

\section{Theory of stellar dynamic stability}
The theory of adiabatic oscillation of gas spheres shows that, 
for homologous radial motion, the square of the angular frequency of the 
lowest radial pulsation mode, in the linear approximation, is given by \citep{led45, led65}:
\begin{equation}
\sigma_{0}^{2} = \frac{\int_{0}^{R} (3\,\Gamma_{1}-4)\, 3\,P_{\rm t}\, dV   }{ \int_{0}^{R} r^{2} \rho \, dV } \, =\,< \Gamma_{1} - \frac{4}{3} > 
\frac{3\, \Omega_{0}}{I_{0}} \,,
\end{equation}  
where $\Omega_{0}$ is the gravitational potential of the star in its equilibrium
state, and $I_{0}$ denotes the moment of inertia with respect to the center of the star.
Since $\Gamma_{1}$ is a function of depth in the star, the integration 
requires geometrical depths down to the stellar center for the computation of $\sigma_{0}^{2}$. 
When $\sigma^{2}_{0}>0$ the configuration is stable, because the standing wave 
solution of the equation of motion does not grow with time.
For supergiants we can assume that the fundamental eigenfrequency $\sigma_{0}$ is not affected 
by magnetic fields, and that rotational kinetic energy can be neglected. 

Stellar dynamic stability depends on the volume-averaged value of $\Gamma_{1}$, and vanishes
when this average 
\begin{equation}
<\Gamma_{1}> \,=\, \frac{\int_{0}^{R}\Gamma_{1}\,P_{\rm t}\,dV }{ \int_{0}^{R}\,P_{\rm t}\,dV } 
\end{equation}
is less than 4/3, hence $\sigma^{2}_{0}\le 0$. Because our detailed knowledge 
of $\Gamma_{1}$ is limited to the atmospheric layers,
it is not straightforward to infer stellar dynamic stability from Eq. (67). 
However, \citet{sto99} recently argued that the integral
can be limited to `an effective basis', or a point where $<$$\Gamma_{1}$$>$
becomes constant in deeper parts of the atmosphere near the base of the stellar envelope. 
Numerical integrations of the equation of motion demonstrate that this truncation is allowed, 
because the radius eigenfunction $\delta$($r$)/$r$ (the relative pulsation amplitude) 
at the base of the outer envelope is already many orders of magnitude smaller than 
its value at the stellar surface (so small that it is virtually indistinguishable from
zero). Since regular stellar pulsation is caused by envelope mechanisms, this limitation of Ledoux' 
stability integral provides a useful means to test for atmospheric stability throughout the HR-diagram. 
Note, however, that a sound analytical foundation for this numerical `criterion' is currently lacking, 
and that important thermodynamic effects due to improved descriptions of $\Gamma_{1}$, as we present here, 
are crucial to ascertain its validity. Nevertheless, atmospheric regions where ionization 
and radiation cause $\Gamma_{1}$ to locally decrease to below 4/3 are of great interest. These real 
gas effects also influence $\Gamma_{2}$ and $\Gamma_{3}$ in supergiants \citep[e.g.][]{lob92}. It is well-known that 
the local lowering of $\Gamma_{2}$ causes the onset of convective motions. \citet{uns48} mentions 
convective `instability'-regions in the partial ionization zones of H and He. For dynamically 
stable atmospheres the local decrease of $\Gamma_{3}$ is linked with an increase or decrease 
of oscillation amplitudes over time, which determines the star's pulsational stability.

\section{Application to supergiant model atmospheres}

\subsection{Model grid}

Based on the NLTE expression for $\Gamma_{1}$ in Sect. 4.2
we investigate dynamic stability of atmospheric models we calculate 
for a range of effective temperatures and gravity accelerations for cool supergiants. 
The new model grid is computed to very high optical depths of log($\tau_{\rm Ross}$)$\simeq$5
with ATLAS9 (R. Kurucz, private communication). A modified version of ATLAS9 is used which, 
for certain models, includes 999 optical depth points. The number of points per 
decade has been improved for the computation of numerical derivatives in the 
treatment of convection. In deep model layers the convection decreases or 
completely vanishes. More information on the ATLAS9 code and these plane-parallel 
hydrostatic models is available from Kurucz\footnote{http://kurucz/grids.html} 
web site. The temperature structure for models with spherical symmetric geometry will 
change significantly, and more detailed solutions of the NLTE-problem (i.e. by 
solving detailed rate equations) will influence the behavior of $\Gamma_{1}$. 
In the present application we assume that the stellar radiation field 
in the optically thin part of the model atmospheres ($\tau_{\rm Ross}$$<$2/3) can 
be approximated by the stellar effective temperature for our computation of 
$\Gamma_{1}$ with height (see Sect. 8 for a discussion).
   
The upper panel of Fig. 3 shows the behavior of $\Gamma_{1}$ ({\em bold solid line}) in the model
with $T_{\rm eff}$=8000~K and log($g$)=1.0.  In the partial NLTE-ionization region of hydrogen 
$\Gamma_{1}$ decreases to $\sim$0.8, around log($\tau_{\rm Ross}$)=$-$2. Towards 
smaller $\tau_{\rm Ross}$, $\Gamma_{1}$ increases and assumes values of $\sim$4/3 ({\em dotted horizontal line}) 
in the outermost atmospheric layers. It results from the ionizing stellar radiation field 
which enhances the mean ionization fraction ({\em solid line in the lower panel}) of very optically thin layers. 
When only collisional ionization is considered ({\em dash-dotted line}), the local kinetic temperature 
becomes too low to appreciably partially ionize the fluid, and $\Gamma_{1}$ 
exceeds unity. The $\Gamma_{1}$-minimum is therefore determined by the incident stellar
radiation field which dilutes above $R_{\star}$ ($\tau_{\rm Ross}$$<$2/3). 
The influence of lowering $W(z)$ in the NLTE-ionization equation and in the radiation pressure 
is shown by the dashed lines for different values of $z$. The dilution of the ionizing radiation 
field diminishes the local value of $\bar{x}$, and the point where $\bar{x}\simeq$0.5 occurs 
in layers of increasingly smaller optical depth. Hence, the corresponding minimum 
in $\Gamma_{1}$ ({\em upper panel}) displaces towards smaller $\tau_{\rm Ross}$-values.     

At large optical depths ($\tau_{\rm Ross}$$>$2/3) the radiation temperature 
approaches the kinetic gas temperature. $\Gamma_{1}$ decreases to below 4/3 due to
thermal (LTE) ionization of hydrogen and helium, whereas a smaller decrease of $\Gamma_{1}$ results from 
$\rm He^{+}$-ionization at the base of the model. At very large optical depths ($\tau_{\rm Ross}$$>$10,000) we  
compute that the fluid becomes fully ionized, and $\Gamma_{1}$ assumes constant values around 1.45. 
Deeper down at the base of the stellar envelope, the value of $\Gamma_{1}$ approaches asymptotically 4/3, 
because radiation pressure outweighs the gas pressure for high temperatures above $10^{5}$~K.

Figure 4 shows the behavior of $\Gamma_{1}$ in model atmospheres for 
4000~K$\leq T_{\rm eff} \leq$20,000~K, with small log($g$)-values. 
The models are plane-parallel and assume a constant value of 2~$\rm km\,s^{-1}$ 
for microturbulence with depth \citep{kur96}. Although the effect of spherical geometry of 
extended atmospheres on the hydrostatic solution for the thermodynamic state is not negligible, 
we presently compare differences in $T_{\rm rad}$ of at least 500~K, which are sufficient to outweigh this
effect. We therefore infer global trends in $\Gamma_{1}$ for a wide range of effective temperatures.   
In the model with $T_{\rm eff}$=4000~K ({\em upper left panel}) $\Gamma_{1}$
assumes values around 5/3 in the outer atmospheric regions. In deeper layers 
$\Gamma_{1}$ lowers to $\simeq$1.15, primarily due to thermal (LTE) ionization of hydrogen. For $T_{\rm eff}$=5000~K
the stellar radiation field becomes sufficiently intense to partially photo-ionize hydrogen at very 
small optical depths, which strongly diminishes $\Gamma_{1}$. We set $W$=1/2 in our further calculations.
When $T_{\rm eff}$ is increased to 6000~K, 7000~K, and 8000~K ({\em upper panel right}),
we find that the partial photo-ionization region displaces towards {\it larger} optical depths.
$\Gamma_{1}$ assumes very small local values $\sim$0.8 for models with $T_{\rm rad}$=7000~K and 
8000~K. This results from the thermal ionization zone of hydrogen, but which displaces towards 
{\it smaller} optical depths for models with higher $T_{\rm eff}$. 
{\it This combined effect whereby the partial thermal ionization zone displaces outwards,
and the photo-ionization region displaces inwards by raising $T_{\rm eff}$,
causes $\Gamma_{1}$ to assume very small values (below unity) over a large geometrical fraction of 
supergiant models with 7000~K$\leq$$T_{\rm eff}$$\leq$8000~K}.

In models with 9000~K$\leq T_{\rm eff} \leq$13,000~K ({\em lower left panel}) we find that the minima 
in $\Gamma_{1}$ increase above $R_{\star}$. This is because hydrogen becomes further ionized, 
or $\bar{x}$$>$0.5 towards higher $T_{\rm rad}$. For the model with $T_{\rm eff}$=11,000~K, hydrogen becomes 
nearly fully ionized, and $\Gamma_{1}$ exceeds 4/3 in the upper atmosphere. In these models the thermal 
He-ionization region occurs at increasingly smaller optical depths, and for $T_{\rm eff}$=13,000~K 
the region approaches $\tau_{\rm Ross}$=2/3 ($\sim$$R_{\star}$). Around these effective (or radiation) temperatures, the partial 
photo-ionization zone of helium enhances in the upper atmospheric layers, and becomes noticeable
by the decrease of $\Gamma_{1}$ to below 4/3. The lower right-hand panel of Fig.~4 shows 
the model with $T_{\rm eff}$=15,000~K, in which the thermal and photo-ionization regions of helium 
merge, which reduces $\Gamma_{1}$ to $\sim$1.25 around $R_{\star}$. We find that atmospheric 
regions with $\Gamma_{1}$ below 4/3 are absent in models with $T_{\rm eff}>$ 16,000~K. 
For the latter, the partial ionization of $\rm He^{+}$ causes a minor decline in $\Gamma_{1}$,
but which exceeds 4/3 throughout the entire model atmosphere.    

\subsection{Model gravity dependence}

A comparison of $\Gamma_{1}$ for models with extended atmospheres  
reveals that their dynamic stability, according Eq. (67), strongly 
depends on $T_{\rm eff}$, or the ionizing stellar radiation field. 
However, differences in $\Gamma_{1}$ with depth in these models also depend on the gravity acceleration. 
Figure 5 shows the contour-map of $\Gamma_{1}$ in the (log($P_{\rm g}$), $T_{\rm e}$)-plane.
We compute $\Gamma_{1}$ with $T_{\rm rad}$=$T_{\rm eff}$,
for models with 6000~K ({\em upper panel}), 8000~K ({\em middle panel}), and 11,000~K ({\em lower panel}), by setting $W$=1/2. 
For $T_{\rm eff}$=6000~K, and $P_{\rm g}$ between 0.1 and 1~dyn~$\rm cm^{-2}$, 
the $\Gamma_{1}$-minima occur for 2000~K$<$ $T_{\rm e}$ $<$8000~K. 
This region marks the upper atmospheric layers where $-$6 $\leq$ log($\tau_{\rm Ross}$) $\leq$$-$4
in the model with log($g$)=1.0 (see {\em upper right-hand panel} of Fig.~4).
The solid diamonds in the upper panel of Fig.~5 show the (log($P_{\rm g}$), $T_{\rm e}$)-values 
of this model, whereas the filled triangles and boxes are plotted for models with 
log($g$)=0.5 and 0.0, respectively. The plots reveal that the $\Gamma_{1}$-values
in layers of small optical depth and $T_{\rm e}$$\leq$4000~K, are not dependent of 
log($g$), and assume comparable values determined by the local gas pressure. 
However, in layers with $\tau_{\rm Ross}$$>$2/3, where $T_{\rm e}$$>$$T_{\rm eff}$=6000~K, 
$\Gamma_{1}$ assumes very different values ranging between 1.6 and 1.36, but which decrease towards
smaller log($g$)-values.      
The same trend is noticeable in the $\Gamma_{1}$ contour-map for $T_{\rm eff}$=8000~K.          
The solid diamonds are shown for log($g$)=2.0, the triangles for 1.5, and the boxes for 1.0.
Around $\tau_{\rm Ross}$=2/3 (where $T_{\rm e}$$\simeq$$T_{\rm eff}$) $\Gamma_{1}$ decreases to below 
unity towards smaller log($g$)-values, as a result of enhanced partial photo-ionization of hydrogen. 
It also reveals that models computed for lower log($g$)-values would assume even smaller 
$\Gamma_{1}$-values in the optically thin region of the atmosphere. However, for log($g$)-values below 1.0 we 
could not converge models to a hydrostatically stable solution with $T_{\rm eff}$=8000~K.
This results from the outwards directed radiation pressure which strongly increases the 
atmospheric density scale-height with smaller gravity acceleration, and which reduces $<$$\Gamma_{1}$$>$ to below 4/3. 
In the next section we show that the $<$$\Gamma_{1}$$>$-integral yields the
atmospheric gravity values for which hydrostatic models of a given $T_{\rm eff}$ become unstable.  

The lower panel of Fig. 5 shows the $\Gamma_{1}$ contour-map for $T_{\rm eff}$=11,000~K.   
The partial hydrogen ionization region, with small $\Gamma_{1}$-values, occurs 
at high kinetic pressures of log($P_{\rm g}$)$\simeq$1.5, because of the   
increased radiation temperature. The decrease of $\Gamma_{1}$ in the partial ionization region of 
helium is also noticeable at these pressures around $T_{\rm e}$$\simeq$16,000~K.      
Three atmospheric models are plotted for log($g$)=3.0
(diamonds), 2.5 (triangles), and 2.2 (boxes). However, the latter has been extrapolated from the 
former two models. The detailed contour map reveals 
that the extrapolation does not lower $\Gamma_{1}$ 
to appreciably smaller values (i.e. to below unity) in the atmosphere. It results from hydrogen 
becoming nearly fully ionized by the incident radiation field,
whereby $\Gamma_{1}$ assumes values around 4/3, also shown in the lower left panel of 
Fig.~4.    
 
\subsection{Stability of model atmospheres}
We evaluate $<$$\Gamma_{1}$$>$ for our grid of model atmospheres.
With \( dV = -d\rho/\rho^{2} \), Eq. (67) transforms to:
\begin{equation}
<\Gamma_{1}> \,=\, \frac{\int_{\rho_{\rm R}}^{\rho_{\star}} \Gamma_{1} \frac{P_{\rm t}}{\rho}\, d\,{\rm ln}\,\rho}
{\int_{\rho_{\rm R}}^{\rho_{\star}} \frac{P_{\rm t}}{\rho}\, d\,{\rm ln}\,\rho } \,,
\end{equation}
where $\rho_{\rm R}$ is the density of the outermost atmospheric layer, 
and $\rho_{\star}$ is the density at the base of the stellar envelope.
In layers of very high optical depth, beyond the partial H- and He-ionization zones, 
$\Gamma_{1}$ becomes locally constant, and $<$$\Gamma_{1}$$>$ assumes an almost constant value. 
The models are dynamically stable against radial mechanic perturbations when 
$<$$\Gamma_{1}$$>$ exceeds 4/3. In general, we find that this condition is valid for 
models which we can converge to a stable solution. However, 
towards smaller log($g$), $<$$\Gamma_{1}$$>$ can decrease to values very close to 4/3. 

Figure 6 shows the run of $\Gamma_{1}$ with log($\tau_{\rm Ross}$) ({\em thin solid lines}).
The corresponding changes in $<$$\Gamma_{1}$$>$ ({\em bold solid lines}) are shown by integrating over 
the atmosphere with Eq. (68), to the density at this optical depth. 
The atmospheric density structure of these cool supergiants is also shown 
for different log($g$)-values of 0.5 ({\em solid lines}), 1.0 ({\em short dash-dotted}), 
1.5 ({\em long dashed}), and 2.0 ({\em long dash-dotted}). For these small gravity accelerations density inversions 
occur by the increase of continuum opacity in the hydrogen ionization region, which 
causes the outward directed radiative pressure gradient to exceed the gravitational 
pull. The condition of hydrostatic equilibrium therefore requires an outward increasing kinetic 
pressure \citep[for a review see][]{mae92}. Pressure inversions occur in our models with 
5250~K$\leq$ $T_{\rm eff}$ $\leq$8250~K, and 0.0$\leq$ log($g$) $\leq$1.5.
The density-inversion layers occur at smaller optical depth with increasing 
$T_{\rm eff}$, because the thermal hydrogen ionization region displaces outwards. 

Our calculations demonstrate that NLTE-ionization of hydrogen in the upper 
atmospheric layers of cool and extended atmospheres strongly diminishes $<$$\Gamma_{1}$$>$
to below 4/3. The value of $<$$\Gamma_{1}$$>$ at a given depth point in Fig. 6 is a measure for 
the dynamic stability of the entire atmosphere above this point. A comparison  
of atmospheric conditions for different $T_{\rm eff}$ between 5500~K and 8500~K reveals that 
changes in the radiation temperature by 1000~K strongly influence the behavior of 
$<$$\Gamma_{1}$$>$ with depth. The partial NLTE-ionization region of hydrogen displaces to 
higher depths by raising $T_{\rm eff}$ from 5500 K to 6500 K ({\em upper panels}).  
The minimum of $\Gamma_{1}$ moves deeper in the atmosphere, and consequently, $<$$\Gamma_{1}$$>$ 
assumes smaller values around these depths. An integration of our model with $T_{\rm eff}$=6500 K
and log($g$)=0.5 to higher depths yields a strong decrease of $<$$\Gamma_{1}$$>$ to below 
unity around log($\tau_{\rm Ross}$)=1.5 ({\em upper panel right}). It partly results from the 
density-inversion which occurs around these depths. The stability integral Eq. (68) is determined 
by the local behavior of $\Gamma_{1}$ {\em and} the atmospheric pressure- and density-structure. 
Towards larger optical depths $<$$\Gamma_{1}$$>$ increases rapidly, because $\Gamma_{1}$ assumes 
local values above 4/3 ({\em dotted horizontal line}), and the density increases steeply. Above $\tau_{\rm Ross}$$\simeq$1000, 
the stability integral assumes almost constant values of $\simeq$1.37 at the base of the stellar envelope.
The integration reveals that this supergiant model, with very low gravity acceleration,
is dynamically stable because the deeper evelope strongly contributes to the overall stability integral.       
However, we could not converge a model with $T_{\rm eff}$=6500 K and a smaller log($g$) of 0.3 
to a stable solution. Models with very small gravity yield $<$$\Gamma_{1}$$>$-values 
below 4/3 when integrating Eq. (68) to the base of the stellar envelope. 

Towards higher $T_{\rm eff}$ of 7500 K ({\em lower left panel}) we compute that the 
minimum of $\Gamma_{1}$ due to NLTE-ionization displaces further down the atmosphere,    
whereas the thermal H- and He-ionization regions occur at lower densities, higher in the atmosphere. 
However, in this model of higher $T_{\rm eff}$, but for the same log($g$)=0.5,
the gas density and pressure also diminish. Therefore, the inward integration of the 
local $\Gamma_{1}$ causes a smaller decrease for $<$$\Gamma_{1}$$>$ at comparable optical depths, 
although $\Gamma_{1}$ assumes locally very small minimum values of $\sim$0.8. On the other 
hand, we find that by integrating this model of increased $T_{\rm rad}$ to very high depths at the base 
of the envelope yields a $<$$\Gamma_{1}$$>$-value of 1.34, or only marginally above 4/3.     
Stable models with log($g$) below 0.5 could not be converged because $<$$\Gamma_{1}$$>$ falls 
to below 4/3 in the deepest layers. In general, we find that models with 
increased $T_{\rm eff}$ (or $T_{\rm rad}$) become unstable towards higher gravity-values
because $<$$\Gamma_{1}$$>$ drops to below 4/3. For $T_{\rm eff}$=8000~K
a stable solution is possible when only log($g$)$\geq$1.0 ({\em lower panel right}), whereas 
for $T_{\rm eff}$=5500~K we can converge stable models for log($g$)$\geq$0.0. 

\subsection{Atmospheric instability regions in cool supergiants}

Stable models of the same gravity yield towards higher $T_{\rm eff}$ smaller $<$$\Gamma_{1}$$>$-values 
at the base of the atmosphere, which is noticeable by comparing the 
four panels of Fig. 6. For a given $T_{\rm eff}$ the stability integral increases with 
log($g$), which demonstrates that the deeper regions of compacter models have a strong
influence by stabilizing the overlying atmosphere. Towards very high log($g$)-values the 
gravity pull effectively balances the radiation pressure gradient, and the density-inversion region 
vanishes. However, our NLTE-calculations of $<$$\Gamma_{1}$$>$ reveal a remarkable  
property. $<$$\Gamma_{1}$$>$ assumes very small values, around unity, down to the base of the 
atmosphere (up to $\tau_{\rm Ross}$$\simeq$100) for models with $T_{\rm eff}$ around 6500 K $-$ 7500 K, 
practically independent of log($g$). Models with higher or lower $T_{\rm eff}$ (of the same gravity), 
yield {\em larger} $<$$\Gamma_{1}$$>$-values at these depths, or they have more stable atmospheres.  

To compare the run of $\Gamma_{1}$ and $<$$\Gamma_{1}$$>$ in models of different $T_{\rm eff}$ and gravity, 
we plot the density scale in Fig. 7, rather than optical depths. The thin drawn
lines show $\Gamma_{1}$, with the deep minimum due to NLTE-ionization of hydrogen, whereas the smaller 
minima at higher densities are due to the LTE-ionization zones. The curious loops in this scale result from the 
density-inversion for ln($\rho$) between $-$24 and $-$20. The boldly drawn 
lines show the corresponding depression in  $<$$\Gamma_{1}$$>$ for models with log($g$)=0.5 ({\em solid lines})
and log($g$)=1.0 ({\em broken lines}). For the latter, $<$$\Gamma_{1}$$>$ assumes the smallest minimum of $\simeq$1.1 
for ln($\rho$)$\simeq$$-$24, around $T_{\rm eff}$=6500 K. In contrast, the value of 
$<$$\Gamma_{1}$$>$ at the base of the model exceeds 4/3 (ln($\rho$)$>$$-$19), but it decreases 
steadily by further raising $T_{\rm eff}$ in steps of 500~K. In other words, the base of these supergiant 
models becomes less stable towards $T_{\rm eff}$ above 6500 K, whereas their extended atmospheric portions
at lower densities tend to become more stable. The latter is also true for models with $T_{\rm eff}$ 
below 6500 K, but instead the deeper envelope further stabilizes the entire model. 

Similar trends occur for models with log($g$)=0.5. We have incorporated the stronger dilution of radiation 
in these more extended atmospheres by setting $z$=1.5, or the geometric dilution factor $W$=(3$-$$\sqrt{5}$)/6 (with Eq. 5)
instead of 1/2, for the calculation of $\Gamma_{1}$ above $R_{\star}$. The minimum in $\Gamma_{1}$  
therefore displaces towards smaller depths (see Sect. 7.1), because partial ionization of hydrogen 
due to the more diluted radiation field occurs at lower densities. Consequently, $<$$\Gamma_{1}$$>$
assumes smaller values at lower densities. Our calculations 
reveal that for models with smaller gravity the minimum of $<$$\Gamma_{1}$$>$ occurs for lower densities, and vice versa.  
The dilution of radiation tends to further destabilize a more extended atmosphere above $R_{\star}$. 
In deeper layers where $T_{\rm rad}$=$T_{\rm e}$, the atmosphere stabilizes because 
$<$$\Gamma_{1}$$>$ increases, but assumes a smaller value above 4/3 (at the base of the envelope)
than for models with higher gravity. Lower gravity models are less stable, and the regions which 
contribute to their destabilization occur at smaller densities in the atmosphere. These 
regions contribute strongest in supergiants with 6500 K$\leq$ $T_{\rm eff}$ $\leq$7500 K, 
as the result of partial non-LTE ionization of hydrogen with $T_{\rm rad}$$\simeq$$T_{\rm eff}$ 
and the partial LTE-ionization of hydrogen and helium, combined with the complex mean density- and 
pressure-structure of such extended atmospheres.  
 
\section{Discussion} 
In a study of the evolution of luminous blue variable (LBV) stars \citet{stc94} evaluated 
$<$$\Gamma_{1}$$>$ in the outer part of the stellar envelope and found that  
massive stars, evolving off the main sequence, develop dynamically unstable outer layers, 
for which eruptive mass-loss can result. \citet{hud94} pointed out that the low surface temperature 
of 12,000~K at which the first instability occurs in these calculations appears to be underestimated, because 
of the absence of very luminous late-B or A-type stars. We note that LTE calculations 
for our model with $T_{\rm eff}$=12,000 K and log($g$)=2.0 show indeed a
decrease of $\Gamma_{1}$ to below 4/3 over a large fraction of the atmosphere with $\tau_{\rm Ross}$$<$2/3 \citep[see Fig. 5 of][]{lob92}.
Towards higher $T_{\rm eff}$, hydrogen fully thermally ionizes, and hence $\Gamma_{1}$ increases above 4/3.
However, our present NLTE calculations of $\Gamma_{1}$ show that stable supergiant models  
with $T_{\rm eff}$ around 12,000 K have rather stabilizing outer layers (with $\Gamma_{1}$ slightly above 4/3; see Fig. 4) 
because the partial photo-ionization region where $\bar{x}$$\simeq$0.5, occurs deeper.  

We note, however, that we have set the radiation temperature equal to $T_{\rm eff}$, and $W$=1/2 or ($3-\sqrt{5}$)/6  
in our calculations. More accurate computations of the detailed NLTE-ionization balance, in which 
$T_{\rm rad}$ varies with height over the atmosphere, and radiation pressure also dilutes with distance, are required to 
determine the precise boundary parameters where the models become dynamically unstable in the HR-diagram. 
Such calculations require `case studies' of various individual supergiants. These stars can have very fast winds and 
high mass-loss rates, which strongly influences the atmospheric extension. Likewise, we expect that small deviations from an `average' 
radiation temperature (but which is proportional to $T_{\rm eff}$) will occur due to variations in the local opacity 
sources at different atmospheric levels. This $T_{\rm rad}$-dependency should be obtained from accurate observations 
of their spectral energy distributions.                               

Our present calculations, which omit these further refinements, show however a clear trend in which $<$$\Gamma_{1}$$>$ 
assumes minimum values below 4/3 in stable models of supergiants with 6500 K $\leq$ $T_{\rm eff}$ $\leq$ 7500 K. 
These atmospheres exist in the extension of the Cepheid instability strip, which has well-defined borders in the HR-diagram.
In this area, the smallest values for $<$$\Gamma_{1}$$>$ occur down to the base of these atmospheres, practically independent of
the gravity acceleration. It indicates the possible relation between pulsation-variability and dynamic destabilization 
mechanisms, caused by the decrease of $<$$\Gamma_{1}$$>$. \citet{sou96} found with time-dependent  
calculations that the atmospheres of these massive stars become unstable to radial pulsations. They also mention that helium 
ionization dominates as the driving mechanism, and the contribution from the (thermal) hydrogen ionization zone becomes significant 
below 5000 K. Long-term spectroscopic and photometric observations of $\rho$~Cas, with $T_{\rm eff}$=6500~K$-$7250~K and 
log($L_{\star}$/$L_{\odot}$)=5.6$-$5.9, demonstrate that non-radial pulsation modes are excited in these extended atmospheres
\citep{lob94}. In a theoretical study of oscillations in cool stars with log($L_{\star}$/$L_{\odot}$)=5,
\citet{shi81} found that stable non-radial modes with $l$=10 can be excited by the ionization zones. 
When the hydrogen and helium ionization zones are situated too shallowly they cannot excite radial pulsations,
but they can drive non-radial modes, which are trapped near the stellar surface. For dynamically 
stable models, pulsation driving occurs in regions where $\Gamma_{3}$ approaches unity. 
\citet{lob92} demonstrated that for conditions of cool supergiant atmospheres, the decrease of $\Gamma_{3}$ 
strongly couples with the decrease of $\Gamma_{1}$ in the partial ionization zones. These considerations
imply that the minima we compute for $\Gamma_{1}$ and $<$$\Gamma_{1}$$>$, for low-gravity stars in the extension 
of the Cepheid strip, can be related with the excitation of stable non-radial pulsations by partial NLTE-ionization 
of hydrogen in the optically thin part of the atmosphere. 

Furthermore, \citet{shi81} mention convectively unstable zones located just below the photosphere 
due to hydrogen ionization in their model with $T_{\rm eff}$=7080 K. In the model with $T_{\rm eff}$=8000 K 
the ionization region emerges above the photosphere, which causes a strong difference between the modal properties 
of both models. In hydrodynamic simulations of compressible convection, which consider the effects of
partial (LTE-) ionization of hydrogen, \citet{ras92} mentions the importance of coupling of convection with pulsations 
in Cepheids. From an observational point of view, we note that photospheric absorption lines of 
yellow hypergiants display an unusually large macrobroadening, which cannot be attributed to rapid rotation (i.e. large 
$v$sin$i$-values) for these evolved and very extended stars. For example, line profile modeling of 
high-dispersion observations in $\rho$~Cas yield macro-broadening velocities of $\sim$25~$\rm km\,s^{-1}$ 
\citep[see Fig. 3 of][]{lob98}. Fast large-scale velocities are possibly linked with ionization-induced formation 
of supersonic vertical flows (plumes), simulated for cool low-luminosity stars \citep[e.g.][]{rat93}.  

Another remarkable aspect of yellow hypergiants \citep[for a review see][]{dej98}
are `eruptions', which occur on time-scales much longer than 
the pulsation quasi-period (ca. half a century, say). In an outburst of 1945-46 $\rho$~Cas (F8p) suddenly dimmed and displayed TiO-bands 
in its spectrum, characteristic of the photospheric temperatures of M-type stars. Within a couple of years (April '47) the star brightened 
up by nearly a magnitude, and a mid G-type spectrum was recovered around 1950. In 1985-86 the star showed a larger-than-average amplitude 
in the light curve, which could be associated 
with shell ejection events \citep{zol91}. Pulsational driving with the occurrence of strong convective motions may excite unstable 
modes with very fast growth rates, resulting in episodic mass ejection. If such mechanism can account for       
eruptions of yellow hypergiants, we expect that it is substantially different for the eruptions of LBVs, 
because strong convective motions induced by partial hydrogen ionization are not expected for these stars, and we compute 
$<$$\Gamma_{1}$$>$-values above 4/3 over the entire atmosphere of hydrostatically stable models with $T_{\rm eff}$$>$ 16,000~K. 

In a series of papers \citet{nie95}, \citet{dej97}, and \citet{nie00}
presented strong theoretical and observational 
indications for the existence of a region in the upper HR-diagram where the atmosphere of 
yellow hypergiants become unstable. The cool boundary of the
`Yellow Evolutionary Void' occurs for {\em bluewards} evolving massive supergiants at $T_{\rm eff}$$\simeq$8300~K and 
log($L_{\star}$/$L_{\odot}$)$>$5.6. Evolutionary calculations show that these stars are expected to evolve along 
tracks of nearly constant luminosity, below the Humphreys-Davidson limit. They evolve bluewards because massive 
stars shed copious amounts of mass to the circumstellar/interstellar environment in the red supergiant phase. 
During redward evolution the very high mass-loss rates reduce the stability of the convective layer, 
and below a critical envelope mass it contracts into a thinner radiative envelope which causes 
rapid blueward evolution. Only for a limited range of initial masses stars 
become red supergiants and evolve back far bluewards. A possible candidate for such a scenario is HR 8752, for which 
\citet{isr99} found an increase of the photospheric temperature by 3000 K$-$4000 K,
based on high-resolution spectra collected over the past 30 years. It is suggested that recurrent eruptions in
yellow hypergiants occur when these stars approach the cool boundary of the Void, and `bounce off' redwards.
The bouncing against the Void may also explain why most of the cool luminous hypergiants cluster near its low-temperature
boundary, while the identification of hypergiants of later spectral type, possibly like VY~CMa (of M-type), is seldom.

In their analysis of atmospheric acceleration mechanisms \citet{nie95}
show that stable solutions cannot be computed for the atmospheres of blueward evolving yellow hypergiants 
at the low-temperature boundary of the Void. It results from a time-independent solution of the
momentum equation, which considers the Newtonian gravity acceleration derived from the evolutionary mass, 
the gas, radiation and turbulent pressure gradients, and the momentum of the stellar wind. 
The latter requires the observed mass-loss rate. An `effective acceleration' for the atmosphere 
is obtained iteratively, but which becomes negative at the cool border of the Void. The outwards directed 
net force causes an unstable atmosphere around $T_{\rm eff}$=8300~K. We note that their 
momentum equation incorporates the important compressibility effects due to thermal ionization 
by the decrease of $\Gamma_{1}$ in the adiabatic sound velocity,
and of the isothermal and isochoric factors ($\chi_{\rho}$ and $\chi_{\rm T}$) in the gas pressure gradient. 
We suggest that the determination of instability boundaries for the Void
be further improved with the NLTE-calculations of $\Gamma_{1}$ we present.
It is important to note that in our computations the minimum of $<$$\Gamma_{1}$$>$ for stable models with lowest gravity 
acceleration occurs around $T_{\rm eff}$=6500~K, which does not coincide with the low-temperature 
border of the Void. However, there is an overlap because $<$$\Gamma_{1}$$>$ approaches values 
very close to 4/3 in low-gravity models at this border. It suggests that the minima we compute for $<$$\Gamma_{1}$$>$
are rather linked with the driving of stable non-radial pulsation modes for the atmospheres of yellow supergiants in the 
extension of the Cepheid instability strip, whereas the Void can result from a particular combination of various acceleration mechanisms, 
combined with the decrease of the overall stability by the larger atmospheric compressibility due to partial NLTE- and LTE-ionization.     
The high mass-loss rates observed for these streaming atmospheres are also determined by 
the sonic point, which is situated inside the photospheres of cool massive supergiants. 
The dynamics of such extended atmospheres must therefore be linked with a prominent mechanism which 
drives these atmospheric pulsations, and which provides the momentum for their supersonic winds.       
 
Finally, we remark that more definitive conclusions about atmospheric variability mechanisms should be obtained
with fully time-dependent hydrodynamic simulations for a number of prototypic hypergiants, like $\rho$~Cas.
Our analytical approximations are useful to evaluate important hydrodynamic quantities as the 
adiabatic sound speed \( v_{\rm ad}^{2} = \Gamma_{1}\,P_{\rm t} / \rho \) in conditions of NLTE, without 
having to numerically solve for large systems of non-linear balance equations. 
With this analytical work we also provide a well-founded basis for more sophisticated numerical 
hydrodynamic calculations. Time-dependent codes which simulate the pulsation of extended 
atmospheres \citep[i.e. described by][]{bes96, sas95} could 
adequately be updated for hydrodynamic NLTE effects, without the loss of their numerical efficiencies. 
With the present study we demonstrate that such advanced calculations cannot be based on the usual 
assumption of local thermodynamic equilibrium. In extended atmospheres, the thermodynamics of non-local 
equilibrium is required, and ultimately, these calculations also ought to consider non-equilibrium 
thermodynamic processes for realistic dynamic modeling.  
      
\section{Conclusions}

1. We present new expressions for the first generalized adiabatic index $\Gamma_{1}$ and the heat capacities,
which are required for the study of dynamic stability of supergiant stellar atmospheres. 
Our equations consider important NLTE effects on the local ionization state by an incident 
and diluted radiation field, not in equilibrium with the local kinetic gas temperature. 
We demonstrate analytically that our more general expressions, which are also valid for multi-component gas mixtures, 
simplify to the classic formulae in which the radiation and kinetic temperature equilibrate.    

2. From a numerical application of our formalism to a grid of supergiant model atmospheres with
solar abundance and 4000~K $\leq$ $T_{\rm eff}$ $\leq$ 20,000~K,
we find that the local values of $\Gamma_{1}$ become very small, below 4/3, over a large fraction 
of the atmosphere for models with $T_{\rm eff}$ between 7000~K and 8000~K. This results from the 
incident radiation field with a temperature of the order of the stellar effective temperature, which 
primarily causes partial photo-ionization of hydrogen. These regions displace deeper down the atmosphere 
with raising $T_{\rm eff}$, whereas the thermal partial ionization regions, at the base of the atmosphere, displace outwards.
This combined effect causes the deep minimum in the local values of $\Gamma_{1}$ to occur 
for these cool star atmospheres. Around these effective temperatures, partial NLTE-ionization causes 
very high atmospheric compressibilities, by which $\Gamma_{1}$ can even decrease to below unity.    

3. Numerical evaluations of Ledoux' stability integral $<$$\Gamma_{1}$$>$ down into the stellar envelope 
demonstrate that $<$$\Gamma_{1}$$>$ exceeds 4/3 in supergiant models for which we can compute 
a hydrostatically stable solution. Towards smaller gravity accelerations $<$$\Gamma_{1}$$>$ decreases, 
and models become unstable with $<$$\Gamma_{1}$$>$ $<$ 4/3 at the base of the envelope. 
Stable models with smaller atmospheric gravity assume smaller $<$$\Gamma_{1}$$>$-values, and
Ledoux' stability integral verifies that models of higher $T_{\rm eff}$  
destabilize at increasingly larger gravity-values due to the enhancement of radiation pressure. 

4. Most importantly, our calculations reveal that for stable supergiant models with 6500~K $\leq$ $T_{\rm eff}$ $\leq$ 7500~K,
$<$$\Gamma_{1}$$>$ assumes minimum values, below 4/3, over a very large fraction of the atmosphere down to its
base. Around $T_{\rm eff}$=6500~K this minimum occurs practically independently of the atmospheric gravity acceleration.
These effective temperatures should be considered valid only within a case study for Kurucz atmospheric models.
For a given $T_{\rm eff}$, in atmospheres of increasingly smaller gravity, $<$$\Gamma_{1}$$>$ assumes small values 
in layers of increasingly lower density due to enhanced partial ionization with the dilution of the radiation field. This corresponds 
with dynamically destabilizing regions, extending over an increasingly larger geometric fraction in the upper atmospheres of more 
massive cool supergiants.  
   
\acknowledgments
I thank Prof. C. de Jager at SRON-Utrecht, for many discussions about 
the physics of supergiant atmospheres which have stimulated the development 
of this theoretical work over the past two years.  Dr. R. Kurucz is gratefully acknowledged for 
calculating the new grid of atmospheric models and helpful discussions. The referee 
is thanked for several useful comments. 
This research is supported in part by an STScI grant GO-5409.02-93A to the 
Smithsonian Astrophysical Observatory.

\clearpage

\begin{figure}
\figcaption[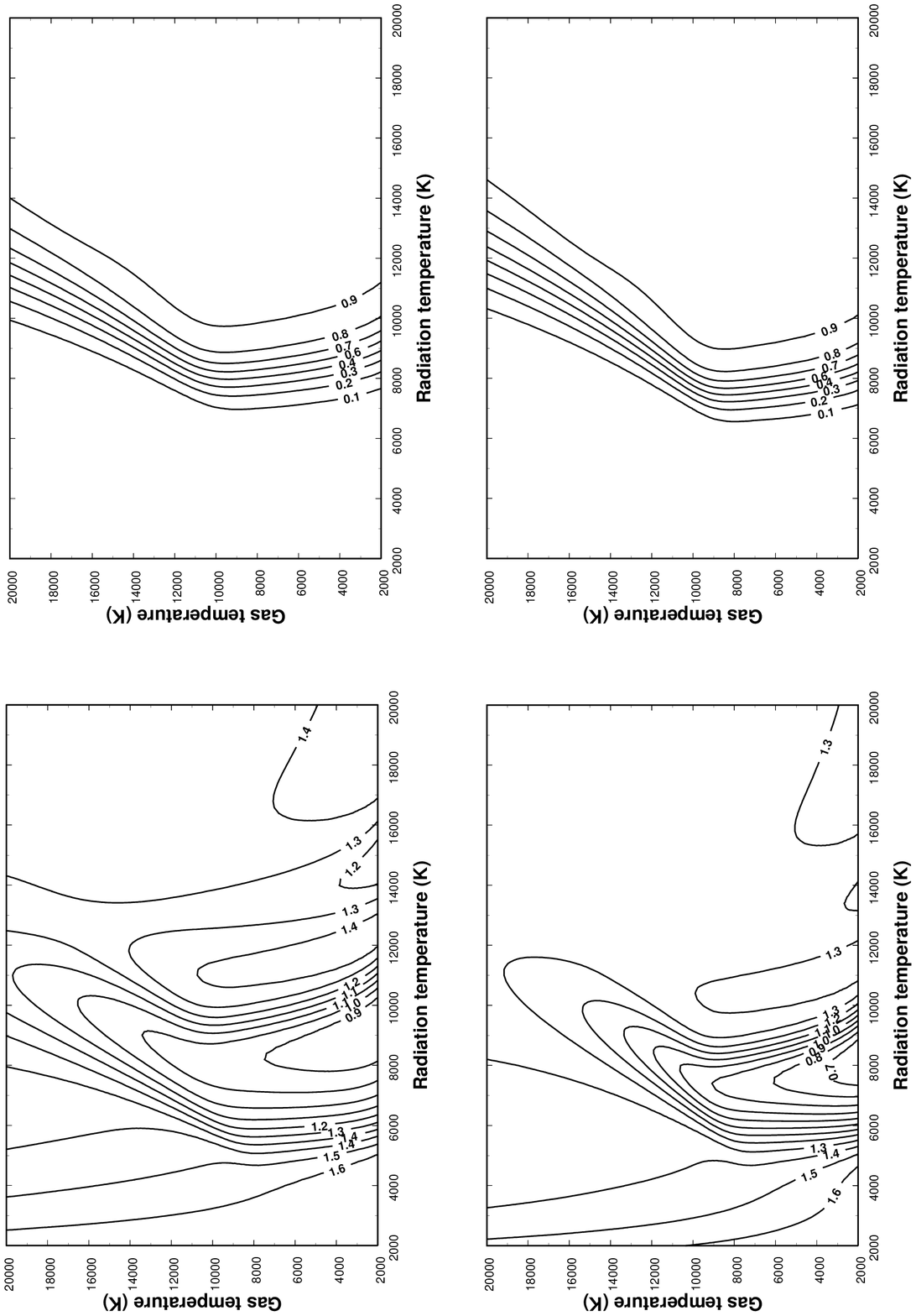]{The left-hand panels show the first adiabatic index $\Gamma_{1}$ computed with partial NLTE-ionization for 
conditions of cool supergiant atmospheres. The mean ionization fraction $\bar{x}$ ({\em panels right}) 
varies with the radiation temperature and the kinetic gas temperature ($T_{\rm e}$). 
$\Gamma_{1}$ assumes minimum values for $\bar{x}$$\simeq$0.5., towards smaller $T_{\rm e}$.
The upper panels are shown for a gas pressure of 10~dyn~$\rm cm^{-2}$, and the lower panels 
for $P_{\rm g}=$0.5~dyn~$\rm cm^{-2}$.\label{fig1}}
\end{figure}
\begin{figure}
\plotone{f1.eps}
\end{figure}

\begin{figure}
\figcaption[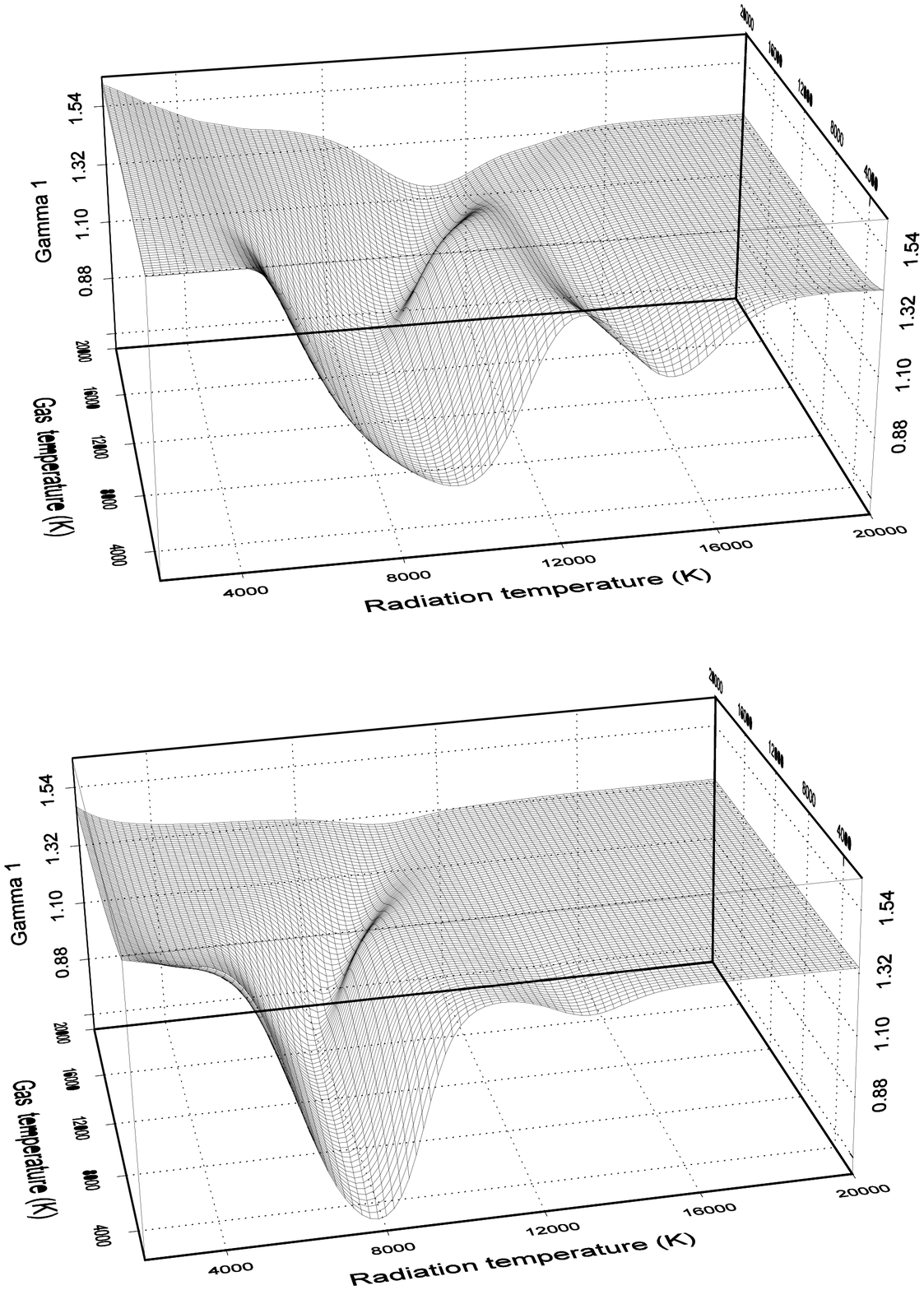]{$\Gamma_{1}$ surface plots for $P_{\rm g}$=10~dyn~$\rm cm^{-2}$ ({\em upper panel})
and $P_{\rm g}$=1~dyn~$\rm cm^{-2}$ ({\em lower panel}). The deep minima of $\Gamma_{1}$ result from partial
NLTE-ionization of hydrogen with 7000~K$\leq$ $T_{\rm rad}$ $\leq$9000 K. The minima between 14,000 K 
and 16,000 K result from the partial ionization of helium. The decrease of $P_{\rm g}$ yields smaller 
minima for $\Gamma_{1}$ with values below 0.7, which corresponds to very high gas compression ratios
for these conditions in the atmospheres of cool supergiants.\label{fig2}}
\end{figure}
\begin{figure}
\plotone{f2.eps}
\end{figure}

\begin{figure}
\figcaption[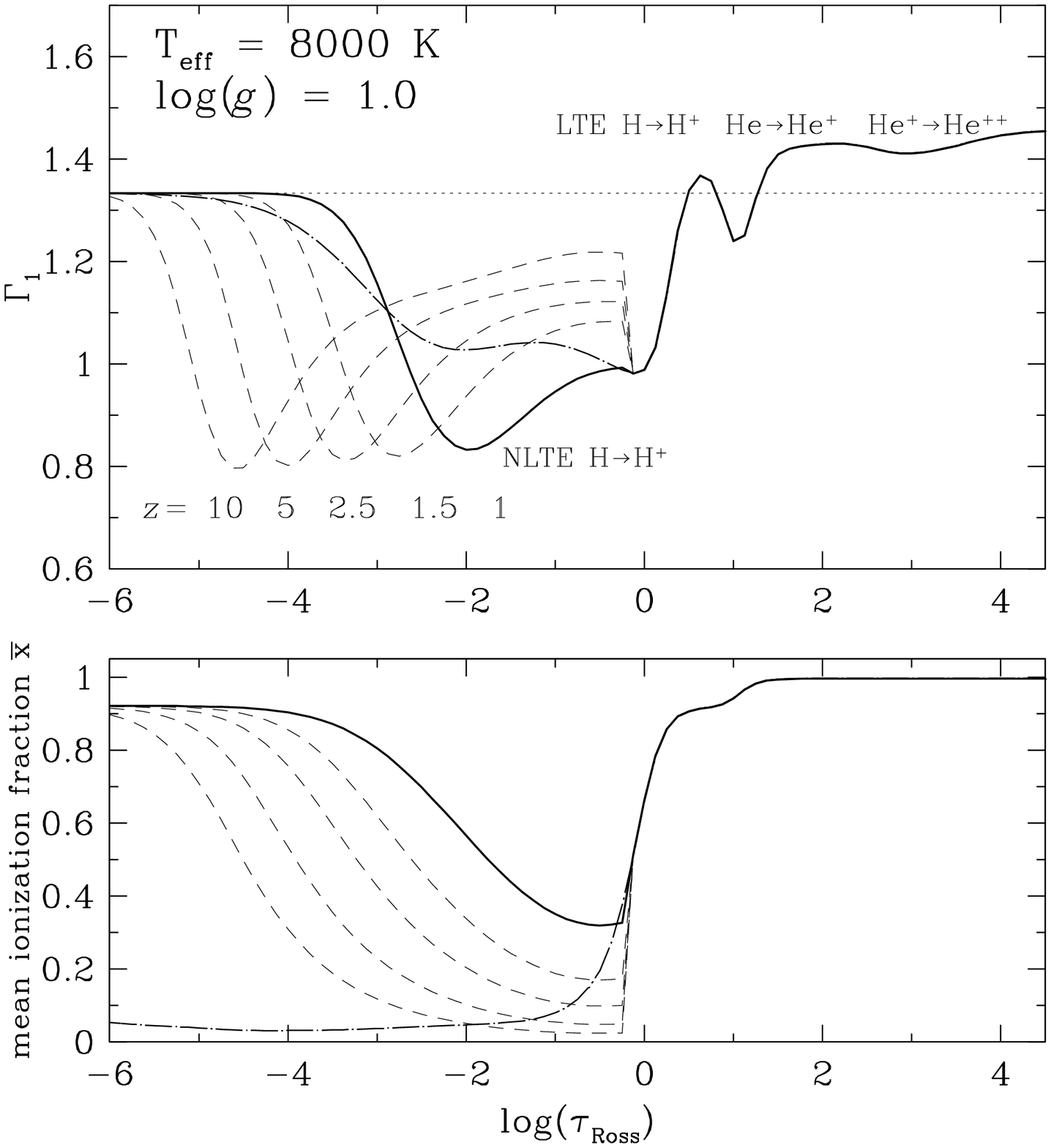]{The run of $\Gamma_{1}$ in the model atmosphere with $T_{\rm eff}$=8000~K and log($g$)=1.0.
The strong decrease of $\Gamma_{1}$ to below 4/3 ({\em dotted horizontal line}) results from partial NLTE-ionization 
for optical depths $\tau_{\rm Ross}$$<$2/3 ($\sim$$R_{\star}$) ({\em bold solid line}).
We assume $T_{\rm rad}$=$T_{\rm eff}$ and $W$=1/2 for the computation of $\Gamma_{1}$ above $R_{\star}$.
At larger depths in the photosphere, $\Gamma_{1}$ mainly decreases due to thermal (LTE) ionization of H and He. The corresponding mean ionization 
fraction $\bar{x}$ is shown in the lower panel ({\em bold solid line}). At smaller optical depths the gas becomes 
fully ionized because $T_{\rm rad}$ exceeds $T_{\rm e}$. For unrealistic conditions of LTE, with
$T_{\rm rad}$=$T_{\rm e}$, these layers would assume a very small $\bar{x}$, with larger $\Gamma_{1}$-values ({\em dash-dotted lines}).       
The influence of the dilution of radiation with distance $z$ above $R_{\star}$ is shown                              
with dashed drawn lines. The mean ionization fraction assumes $\bar{x}$$\simeq$0.5 at smaller optical depths when 
the ionizing radiation field dilutes more with increasing distance $z$. Consequently, the deep minimum of $\Gamma_{1}$ in the upper panel, 
due to partial NLTE-ionization of hydrogen, occurs at smaller optical depths.\label{fig3}}
\end{figure}
\begin{figure}
\plotone{f3.eps}
\end{figure}

\begin{figure}
\figcaption[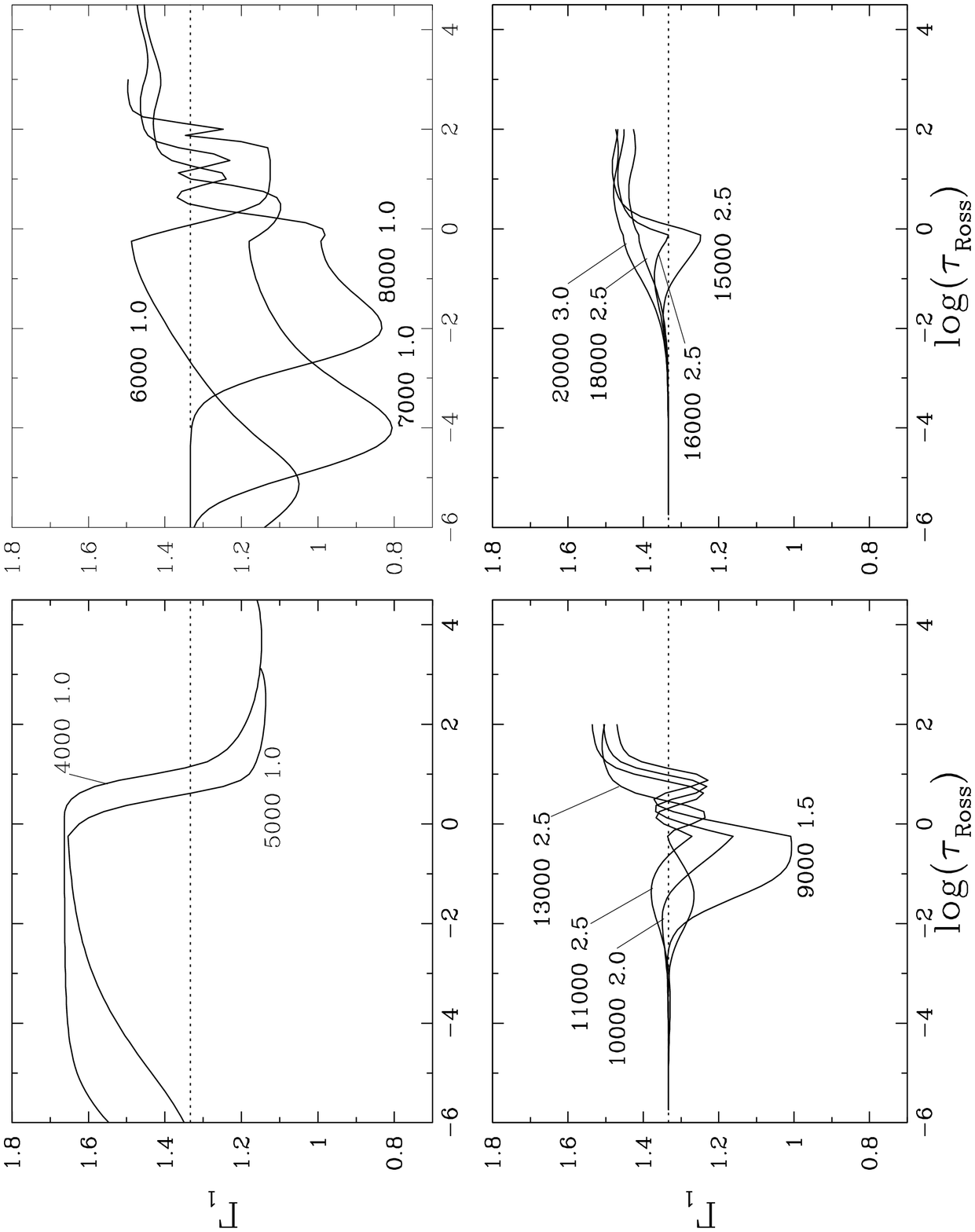]{The behavior of $\Gamma_{1}$ in the outer envelope of supergiants with 
4000~K$\leq$ $T_{\rm eff}$ $\leq$20,000 K. Model $T_{\rm eff}$- and log($g$)-values are labeled. The 
NLTE $\Gamma_{1}$ assumes smallest values of $\sim$0.8 above $R_{\star}$ ($\tau_{\rm Ross}$$<$2/3) for models with $T_{\rm eff}$ between 
7000~K and 8000~K ({\em upper panel right}). 
We assume $T_{\rm rad}$=$T_{\rm eff}$ and $W$=1/2 for the computation of $\Gamma_{1}$ above $R_{\star}$.
Beneath $R_{\star}$, the LTE $\Gamma_{1}$-values 
decrease to below 4/3 ({\em horizontal dotted lines}) due to partial ionization of H and He. 
These thermal ionization zones displace outwards in models with higher $T_{\rm eff}$, while the partial hydrogen NLTE-ionization region in the 
outer atmosphere displaces inwards. This causes the deep $\Gamma_{1}$-minimum for 7000~K$\leq$ $T_{\rm eff}$ $\leq$8000~K.         
$\Gamma_{1}$ assumes increasingly larger values for models of higher $T_{\rm eff}$ ({\em lower panels}). 
For models with $T_{\rm eff}$$\geq$16,000 K and log($g$)$\geq$2.5 helium is nearly fully ionized, and $\Gamma_{1}$ assumes values above 4/3 
over the entire atmosphere (see text).\label{fig4}}
\end{figure}
\begin{figure}
\plotone{f4.eps}
\end{figure}

\begin{figure}
\figcaption[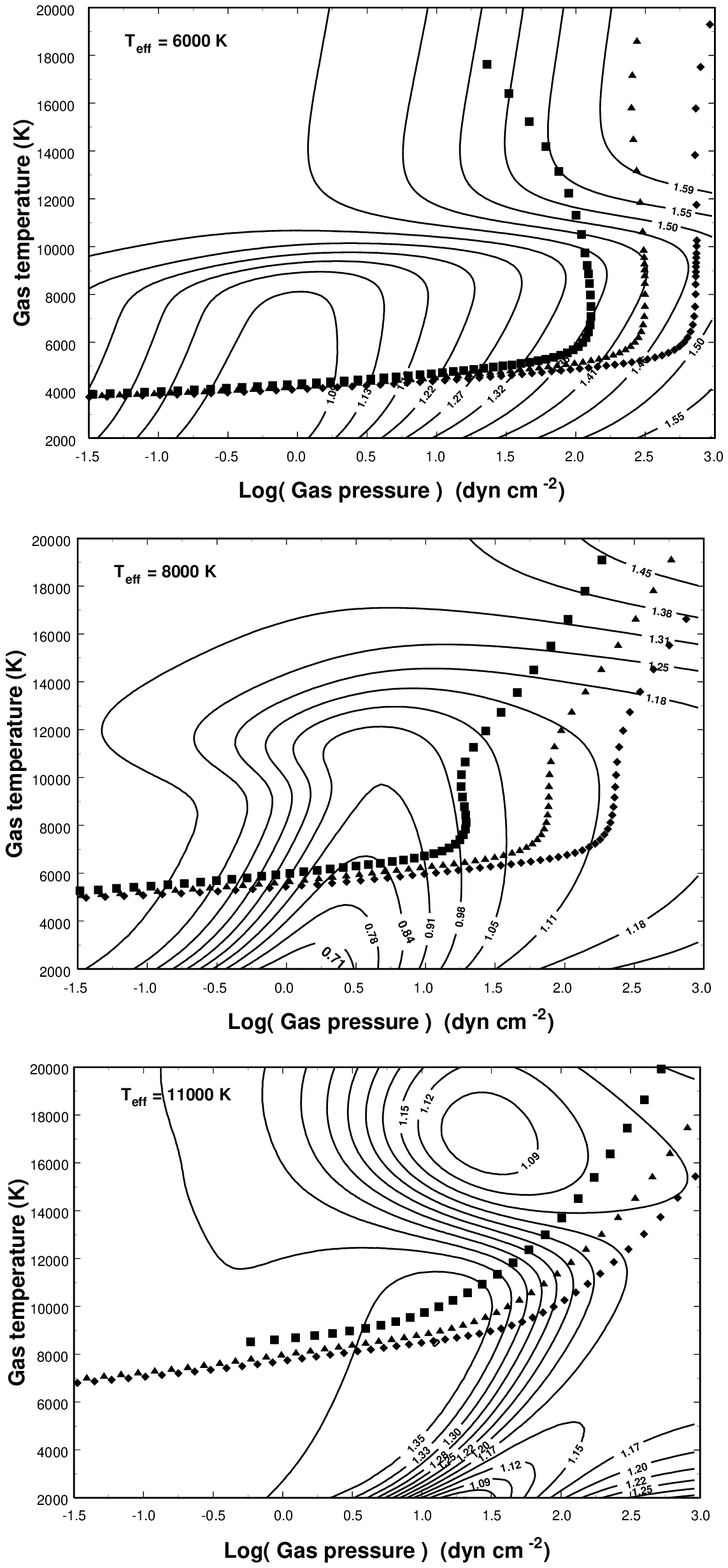]{{\em Upper panel:} Atmospheric models of supergiants with $T_{\rm eff}$=6000~K are plotted in
the (log($P_{\rm g}$), $T_{\rm e}$)-plane, for three gravity accelerations of log($g$)=1.0 ({\em diamonds}), 0.5 ({\em triangles}), and 0.0 ({\em boxes}).
The solid lines show the contour map of $\Gamma_{1}$, computed with NLTE for $T_{\rm rad}$=$T_{\rm eff}$ and $W$=1/2.  
The run of $\Gamma_{1}$ in the outer atmospheric layers, with $T_{\rm e}$$\simeq$4000 K$-$5000~K, is very similar. 
The models differ more in the deeper layers, where different NLTE $\Gamma_{1}$-values are assumed. 
In layers with $T_{\rm e}$$\simeq$$T_{\rm eff}$ (around $R_{\star}$), $\Gamma_{1}$ decreases towards smaller 
log($g$)-values because the lines of equal $\Gamma_{1}$ decrease parallel with the local (log($P_{\rm g}$), $T_{\rm e}$)-structure. 
In deeper layers (below $R_{\star}$), where $T_{\rm e}$$>$$T_{\rm eff}$, $\Gamma_{1}$-values evaluated with NLTE-ionization 
stay above 4/3. In these layers, LTE ionization calculations with $T_{\rm rad}$ set equal to $T_{\rm e}$ are required. \newline
{\em Middle panel:} Three models with $T_{\rm eff}$=8000 K are plotted for log($g$)=2.0 ({\em diamonds}), 1.5 ({\em triangles}), and 1.0 ({\em boxes}).
The NLTE $\Gamma_{1}$ contour map is computed for $T_{\rm rad}$=8000 K. In layers near $R_{\star}$ ($T_{\rm e}$$\simeq$$T_{\rm eff}$), 
$\Gamma_{1}$ assumes values below 4/3. The map reveals that in models with log($g$) below 1.0 these layers will assume even 
smaller $\Gamma_{1}$-values, because their $\Gamma_{1}$-contours run parallel with the (log($P_{\rm g}$), $T_{\rm e}$)-lines. Stable hydrostatic 
solutions cannot be computed for these small gravity accelerations. \newline
{\em Lower panel:} Three models with $T_{\rm eff}$=11,000 K are plotted for log($g$)=3.0 ({\em diamonds}), 2.5 ({\em triangles}), and 2.2 ({\em boxes}). 
In layers near $R_{\star}$ ($T_{\rm e}$$\simeq$$T_{\rm eff}$) the $\Gamma_{1}$ contour map varies perpendicular to the atmospheric structure variations 
with gravity of the models. The layers above $R_{\star}$ assume $\Gamma_{1}$-values around 4/3. 
Since hydrogen becomes nearly fully ionized, the upper atmospheres are more stable compared 
to the models with $T_{\rm eff}$=6000 K and 8000 K.\label{fig5}}
\end{figure}
\begin{figure}
\plotone{f5.eps}
\end{figure}

\begin{figure}
\figcaption[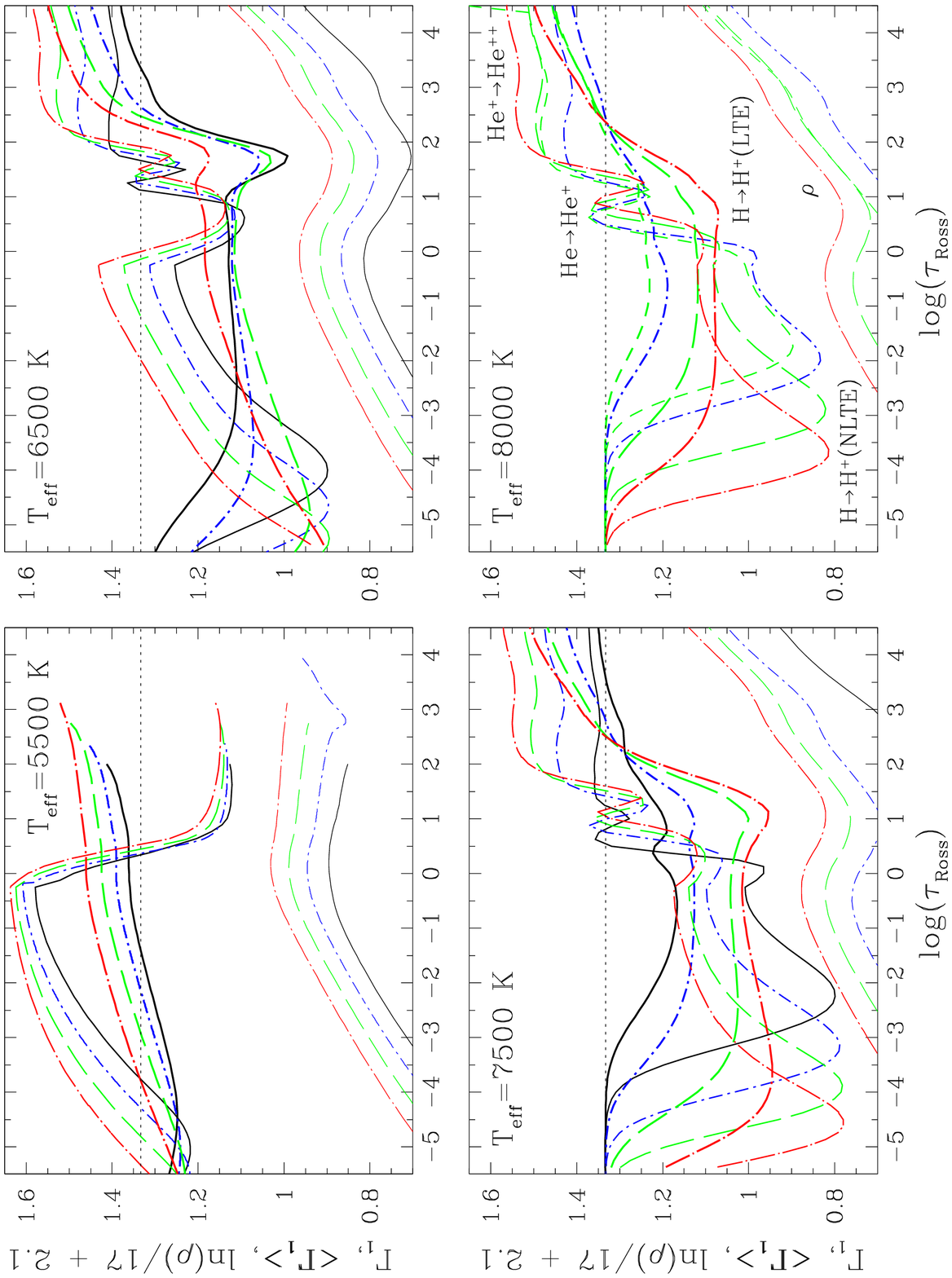]{The behavior of $\Gamma_{1}$ ({\em thin lines}) and $<$$\Gamma_{1}$$>$ ({\em bold solid lines}) with optical depth
in envelope models of cool supergiants for different gravity accelerations. $<$$\Gamma_{1}$$>$ is integrated to the density at the corresponding
optical depth. We assume $T_{\rm rad}$=$T_{\rm eff}$ and $W$=1/2 for the computation of $\Gamma_{1}$ above $R_{\star}$.
Higher gravity models are more stable because $<$$\Gamma_{1}$$>$ increases for a given $T_{\rm eff}$ at the base of the envelope 
with increasing log($g$)=0.5 ({\em solid lines}), 1.0 ({\em short dash-dotted}), 1.5 ({\em long dashed}), and 2.0 ({\em long dash-dotted}). 
For a given log($g$), towards higher $T_{\rm eff}$, $<$$\Gamma_{1}$$>$ decreases at the base of the 
envelope to values close to 4/3 ({\em dotted horizontal line}). Models with log($g$)$<$1.0 are unstable for $T_{\rm eff}$$>$8000~K.
The model with $T_{\rm eff}$=8500 K and log($g$)=1.5 ({\em short dashed lines in lower panel right}) is also shown.
The partial ionization zones of hydrogen and helium, where $\Gamma_{1}$ locally decreases, are labeled.
For $T_{\rm eff}$=6500 K and 7500 K, $<$$\Gamma_{1}$$>$ assumes minimum values down to the base of the atmosphere (log($\tau_{\rm Ross}$)$\sim$1.5)
due to partial NLTE and LTE H- and He-ionization, combined with the density- and pressure-inversion regions of these models. 
Note the near gravity-independence for these minima in the models with $T_{\rm eff}$=6500~K (see text).\label{fig6}}
\end{figure}
\begin{figure}
\plotone{f6.eps}
\end{figure}

\begin{figure}
\figcaption[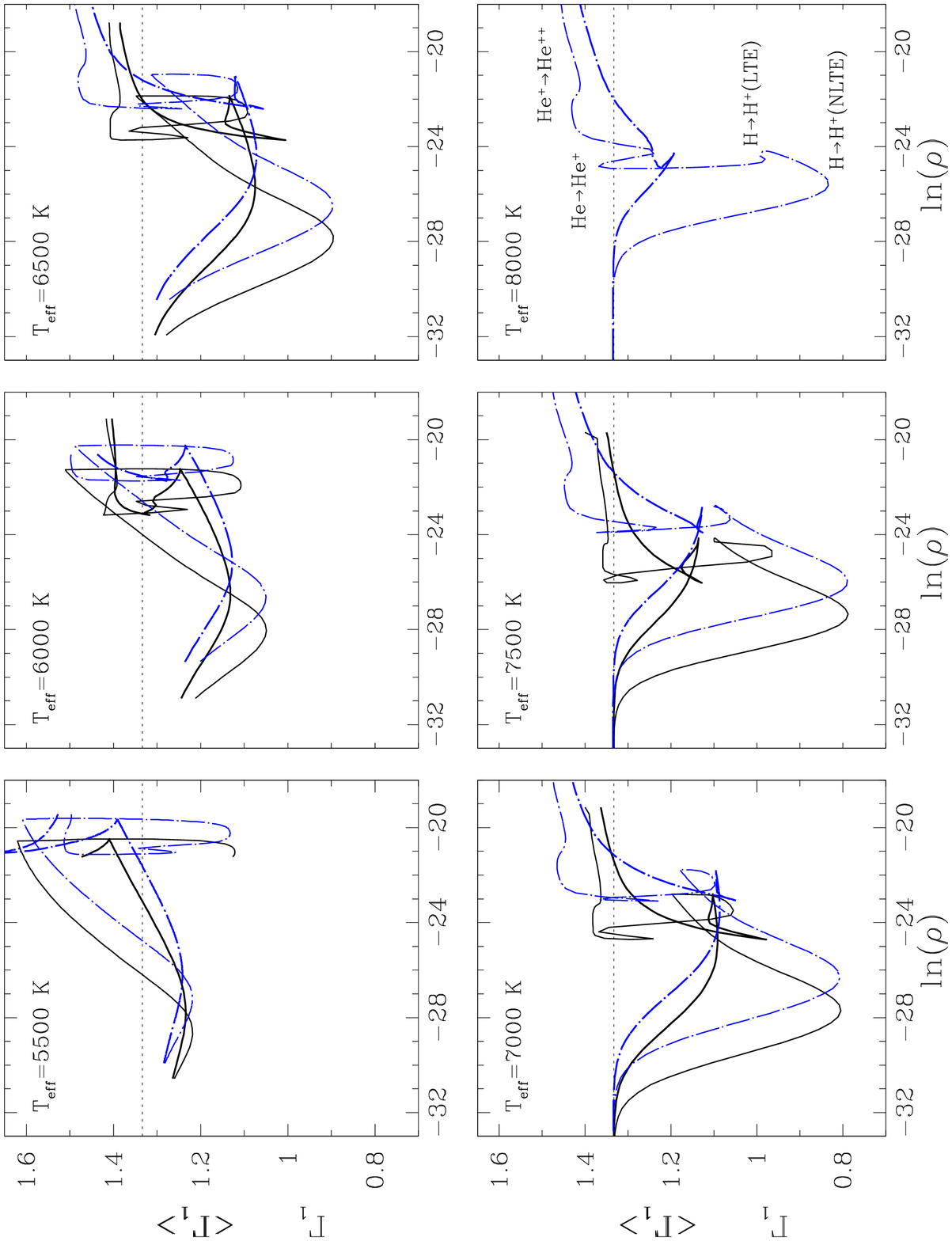]{The run of $\Gamma_{1}$ ({\em thin lines}) and $<$$\Gamma_{1}$$>$ ({\em bold lines}) in the atmospheric density scale 
for log($g$)=0.5 ({\em solid lines}) and 1.0 ({\em broken lines}). $<$$\Gamma_{1}$$>$ assumes minimum values (below 4/3) in the atmosphere
of stable models with 6500 K$\leq$ $T_{\rm eff}$ $\leq$7500 K. We assume $T_{\rm rad}$=$T_{\rm eff}$ for the computation of $\Gamma_{1}$ 
above $R_{\star}$. Towards smaller gravity acceleration the destabilizing regions, 
due to partial NLTE-ionization of hydrogen, occur at lower density with the dilution of the ionizing radiation field (see text).\label{fig7}}
\end{figure} 
\begin{figure}
\plotone{f7.eps}
\end{figure}

\clearpage
\appendix
\centerline{\bf APPENDIX}

The first adiabatic index is defined by the adiabatic thermodynamic
derivative:
\begin{equation}
\Gamma_{1} \equiv \left( \frac{\partial\,{\rm ln}\,P_{\rm t}}{\partial\,{\rm ln}\,\rho} \right)_{\rm ad} \,,
\end{equation}
which can also be expressed by combining three thermodynamic quantities:
\begin{equation}
\Gamma_{1}= \frac{c_{P_{\rm t}}}{c_{v}}\chi_{\rho} \,,
\end{equation}
where \( c_{v} \), \( c_{P_{\rm t}} \) and \( \chi_{\rho} \) denote the
heat capacity at constant volume, at constant total pressure,
and the isothermal factor, respectively.
The total pressure, including the radiation pressure is given by the following equation of state:
\begin{equation}
P_{\rm t} = P_{\rm g} + P_{\rm rad} = N\,k\,T_{\rm e}\,(1+ \bar{x})\,\rho+\frac{1}{3}\,a\,W\,T^{4}_{\rm rad}\,,
\end{equation}
where $k$ is the Boltzmann constant, $N$ the total number of particles 
per unit mass, and $a$ is the radiation density constant.
$W$ is the geometric dilution function for the radiation pressure 
with distance $d$ from the stellar surface $R_{\star}$. At the surface 
$z$=$d$/$R_{\star}$=1, hence $W$($z$=1)=1/2 with  
\begin{equation}
W(z) = 1/2\,(1 - \sqrt{1 - 1/z^{2}}) \,.
\end{equation}
The ionization equation, modified for photo-ionization from the ground
level for every element $i$, is (Sect. 2.1):
\begin{equation}
\frac{x_{i}}{1-x_{i}}\,\bar{x}=C_{i}\,W\, \frac{T_{\rm e}^{\frac{1}{2}}}{\rho}\, T_{\rm rad} \,
\,{\rm exp}  \left( -\frac{I_{i}}{k\,T_{\rm rad}} \right) \,,
\end{equation}
where \( C_{i} \) is a constant, and \( x_{i} \) the ionization fraction of element $i$, which is singly-ionized 
from the ground level, with ionization 
energy \( I_{i} \), by the incident radiation field of temperature 
\( T_{\rm rad} \). We denote the mean ionization fraction of the mixture 
with $m$ elements:
\begin{equation}
\bar{x}= \sum_{i=1}^{m} \nu_{i}\,x_{i} \,,
\end{equation}
where \(\nu_{i}= N_{i}/N \) is the element abundance having $N_{i}$ particles 
per unit mass, and \( \sum_{i} \nu_{i} = 1 \) or \( N = \sum_{i} N_{i} \).
Hence, the total number of electrons per unit mass is \( N_{\rm e} = \bar{x}\,N\).
    
When the radiation temperature $T_{\rm rad}$ and the kinetic gas temperature 
$T_{\rm e}$ are independent quantities, the specific heat capacity (per unit mass) at 
constant volume $v$ (or density, since $v$=1/$\rho$) is defined as the sum 
of the internal energy derivatives to both independent temperatures:
\begin{equation}
c_{v} \equiv \left( \frac{\partial\,e}{\partial \, T_{\rm e}}\right)_{\rho,\,T_{\rm rad}} + \left( \frac{\partial\,e}{\partial \, T_{\rm rad}}\right)_{\rho,\,T_{\rm e}} \,.
\end{equation}
The internal energy function in Eq. (7) is given by:
\begin{equation}
e = N\,k\,T_{\rm e} \left( \frac{3}{2}(1+\bar{x}) + \sum_{i=1}^{m} \nu_{i} \left(
x_{i} \frac{I_{i}}{kT_{\rm e}} + x_{i} \frac{w_{i}}{u_{i}}
+ x^{0}_{i} \frac{w^{0}_{i}}{u^{0}_{i}} \right) \right) + \frac{a\,W\,T_{\rm rad}^{4}}{\rho} \,, 
\end{equation}
where \( x^{0}_{i} = 1 - x_{i} \) denotes the neutral particle fraction.
The two electronic energy terms of Eq. (8)
include the partition functions which are defined
by: \( u_{i} = \sum_{r=0}^{\infty} g_{r,i} \, {\rm e}^{-\frac{\chi_{r,i}}{kT_{\rm e}}} \)
and \( w_{i} = \sum_{r=0}^{\infty} g_{r,i}
\left( \frac{\chi_{r,i}}{kT_{\rm e}} \right) \, {\rm e}^{-\frac{\chi_{r,i}}{kT_{\rm e}}} \),
where \( g_{r,i} \) is the statistical weight of the \( r^{\rm th} \)
excitation level.
The partition functions of the neutral fractions are indicated 
with the (0) superscript. 

In conditions of stellar atmospheres, the kinetic and ionic 
terms largely outweigh the electronic terms, and since we assume ionization 
from the ground level only, the electronic terms can be neglected, yielding:
\begin{equation}
e =  \frac{3}{2}\,N\,k\,T_{\rm e}\,(1+\bar{x}) + N\,\sum_{i=1}^{m} \nu_{i}\, 
x_{i}\,I_{i}  + \frac{a\,W\,T_{\rm rad}^{4}}{\rho} \,. 
\end{equation}

The specific heat capacity at constant total pressure in Eq. (2) is defined 
as the sum of the derivatives of the enthalpy function \( h = e + P_{\rm t} / \rho \) 
to both independent temperatures:
\begin{equation}
c_{P_{\rm t}} \equiv \left( \frac{\partial \, h}{\partial \, T_{\rm e}} \right)_{P_{\rm t},\, T_{\rm rad}}+ \left( \frac{\partial \, h}{\partial \, T_{\rm rad}} \right)_{P_{\rm t},\, T_{\rm e}} \,, 
\end{equation}
where
\begin{equation}
h=\frac{5}{2}N\,k\,T_{\rm e}\,(1+\bar{x})+N\,\sum_{i=1}^{m} \nu_{i}\,
x_{i}\,I_{i} + \frac{\frac{4}{3}\,a\,W\,T_{\rm rad}^{4}}{\rho} \,.
\end{equation}

We proceed by deriving the detailed expressions for the individual terms of Eq. (7) and Eq. (10).

In Eq. (7) the kinetic temperature derivative at constant density yields:
\begin{equation}
\left( \frac{{\partial} \, e}{{\partial} \, T_{\rm e}} \right)_{\rho} = 
\frac{3}{2} N \, k \left( (1+\bar{x}) + T_{\rm e}
\left( \frac{{\partial} \, \bar{x}}{{\partial} \, T_{\rm e}} \right)_{\rho} \right)  
+ N \sum_{i} \nu_{i}\, I_{i} \, \left( \frac{{\partial} \, x_{i}}{{\partial} \, T_{\rm e}} \right)_{\rho} \,,
\end{equation} 
and the radiation temperature derivative yields:
\begin{equation}
\left( \frac{{\partial} \, e}{{\partial} \, T_{\rm rad}} \right)_{\rho} = 
\frac{3}{2} N \, k \, T_{\rm e}
\left( \frac{{\partial} \, \bar{x}}{{\partial} \, T_{\rm rad}} \right)_{\rho}  
+ N \, \sum_{i} \nu_{i}\, I_{i} \, \left( \frac{{\partial} \, x_{i}}{{\partial} \, T_{\rm rad}} \right)_{\rho} + 12\,\alpha\,\theta\,N\,k\,(1+\bar{x}) \,,
\end{equation}
where \( \alpha \) is the ratio of the radiation pressure and the kinetic gas
pressure \( P_{\rm rad}/P\), and \( \theta \) is the ratio of the local
kinetic temperature and the temperature of the incident radiation field \( T_{\rm e}/T_{\rm rad} \). 

In Eq. (10) the kinetic temperature derivative at constant total pressure yields:
\begin{eqnarray}
\left( \frac{\partial \, h}{\partial \, T_{\rm e}} \right)_{P_{\rm t}} = & &
\frac{5}{2}\,N\,k \left( (1+\bar{x}) + T_{\rm e} \left( \frac{\partial \bar{x}}{\partial T_{\rm e}} \right)_{P_{\rm t}} \right)      
+ N\,\sum_{i} \nu_{i}\,I_{i} \left( \frac{\partial \, x_{i}}{\partial \, T_{\rm e}}\right)_{P_{\rm t}} \nonumber \\ 
& & -\,4\alpha\,N\,k\,(1+\bar{x}) \left( \frac{\partial \, {\rm ln}\,\rho}{\partial\,{\rm ln}\,T_{\rm e}} \right)_{P_{\rm t}} \,,
\end{eqnarray}
and the radiation temperature derivative casts by using \( P_{\rm rad} = \frac{1}{3}\,a\,W\,T_{\rm rad}^{4} \) into:
\begin{eqnarray}
\left( \frac{\partial h}{\partial T_{\rm rad}} 
\right)_{P_{\rm t}} = & & 
\frac{5}{2}\,N\,k\,T_{\rm e}\, 
\left( \frac{ \partial \bar{x} }{ \partial T_{\rm rad} } 
\right)_{P_{\rm t}} 
+ N\, \sum_{i} \nu_{i}\,I_{i}\,
\left( \frac{ \partial x_{i}   }{ \partial T_{\rm rad} } 
\right)_{P_{\rm t}} \nonumber \\
& & + 4\,\alpha\,N\,k\,\theta\,(1+\bar{x})
\left( 4 - 
\left( \frac{ \partial \, {\rm ln}\,\rho }{ \partial\,{\rm ln}\,T_{\rm rad} } 
\right)_{P_{\rm t}} 
\right) \,.
\end{eqnarray}

Hence, both heat capacities are obtained from the detailed evaluation of two thermodynamic quantities
in Eq. (12) and Eq. (13):
\begin{equation}
\left( \frac{ \partial x_{i} }{\partial T_{\rm e} }  \right)_{\rho} \,\,, \,\,  {\rm and} \,\, 
\left( \frac{ \partial x_{i} }{\partial T_{\rm rad} }  \right)_{\rho} \,,
\end{equation}
and with Eq. (6)
\begin{equation}
\left( \frac{ \partial \bar{x} }{\partial T_{\rm e} }  \right)_{\rho} = 
\sum_{i} \nu_{i} \, \left( \frac{ \partial x_{i} }{\partial T_{\rm e} }  \right)_{\rho} \,, \,\, {\rm and} \,\,
\left( \frac{ \partial \bar{x} }{\partial T_{\rm rad} }  \right)_{\rho} =
\sum_{i} \nu_{i} \,  \left( \frac{ \partial x_{i} }{\partial T_{\rm rad} }  \right)_{\rho} \,.
\end{equation}
Four thermodynamic quantities are to be evaluated in Eq. (14) and Eq. (15):
\begin{equation}
\left( \frac{ \partial x_{i} }{\partial T_{\rm e}   }  \right)_{P_{\rm t}} \,, 
\left( \frac{ \partial x_{i} }{\partial T_{\rm rad} }  \right)_{P_{\rm t}} \,,
\left( \frac{ \partial \rho  }{\partial T_{\rm e}   }  \right)_{P_{\rm t}} \,, \,\, {\rm and} \,\, 
\left( \frac{ \partial \rho  }{\partial T_{\rm rad} }  \right)_{P_{\rm t}} \,,
\end{equation}
for which also  
\begin{equation}
\left( \frac{ \partial \bar{x} }{\partial T_{\rm e} }  \right)_{P_{\rm t}} = 
\sum_{i} \nu_{i} \, \left( \frac{ \partial x_{i} }{\partial T_{\rm e} }  \right)_{P_{\rm t}} \,\, {\rm and} \,\,
\left( \frac{ \partial \bar{x} }{\partial T_{\rm rad} }  \right)_{P_{\rm t}} =
\sum_{i} \nu_{i} \,  \left( \frac{ \partial x_{i} }{\partial T_{\rm rad} }  \right)_{P_{\rm t}} \,.
\end{equation}

$i$. The kinetic temperature derivative of the ionization equation (5) at constant density and radiation temperature yields
after grouping terms and some rearrangement:
\begin{equation}
 \left( \frac{ \partial x_{i} }{\partial\, {\rm ln}\, T_{\rm e} }  \right)_{\rho} =
x_{i}\,(1-x_{i})\,\left( \frac{1}{2} -   \left( \frac{ \partial\, {\rm ln}\, \bar{x} }{\partial \, {\rm ln} \, T_{\rm e} }  \right)_{\rho} \right) \, 
 \,,
\end{equation}
which casts with Eq. (17) (multiplying both sides with \( \sum_{i} \nu_{i} \) ), after factorization, into:
\begin{equation}
 \left( \frac{ \partial\, \bar{x} }{\partial \, T_{\rm e} }  \right)_{\rho} =
\frac{1}{2\,T_{\rm e}}\,\,\frac{\bar{x}\,\sum_{i} \nu_{i} x_{i} (1-x_{i})}{\sum_{i} \nu_{i} x_{i} (1-x_{i}) + \bar{x} } \,.
\end{equation}
We obtain by inserting Eq. (21) in Eq. (20) for every element $i$:
\begin{equation}
 \left( \frac{ \partial\, x_{i} }{\partial\, T_{\rm e} }  \right)_{\rho} =
\frac{1}{2\,T_{\rm e}}\,\,\frac{x_{i} (1-x_{i})\,\sum_{i}\,\nu_{i}\,x_{i}}{\sum_{i} \nu_{i} x_{i} (1-x_{i}) + \sum_{i}\nu_{i}\,x_{i} } \,.
\end{equation}
Equation (12) reduces, after inserting Eq. (21) and Eq. (22) and  grouping terms, to:
\begin{eqnarray}
\left( \frac{{\partial} \, e}{{\partial} \, T_{\rm e}} \right)_{\rho} & & =
 N\,k\, \left( \frac{3}{2}(1+\bar{x}) + \frac{1}{2}\,  \left( \frac{3}{2} \bar{x} -  
\sum_{i} \nu_{i} x_{i} (1-x_{i}) \frac{I_{i}}{k\,T_{\rm e}} \right) \frac{ \sum_{i}\nu_{i} x_{i} (1-x_{i}) }{\bar{x} + 
\sum_{i} \nu_{i} x_{i} (1-x_{i})  } \right) \nonumber \\
& & + N\,k\, \frac{1}{2} \sum_{i} \nu_{i} x_{i} (1-x_{i}) \frac{I_{i}}{k\,T_{\rm e}}  \\ 
& & = N\,k\, \left( \frac{3}{2}(1+\bar{x}) + \frac{\bar{x}}{2} \, \frac{\sum_{i} \nu_{i} x_{i} (1-x_{i}) \left( \frac{3}{2} 
+ \frac{I_{i}}{k\,T_{\rm e}} \right) }{\bar{x} + \sum_{i} \nu_{i} x_{i} (1-x_{i})  }    \right) \,.
\end{eqnarray} 

$ii$. Analogously, the radiation temperature derivative of the ionization equation (5) at constant density and kinetic temperature yields
after grouping terms and some rearrangement:
\begin{equation}
\left( \frac{\partial x_{i}}{\partial \, {\rm ln} \, T_{\rm rad}} \right)_{\rho} =
x_{i}\,(1-x_{i})\, \left( \left(1+ \frac{I_{i}}{k\, T_{\rm rad}}\right) - \left( \frac{\partial \, {\rm ln} \, \bar{x}}{ \partial \, {\rm ln} \, 
T_{\rm rad} }\right)_{\rho} \right)  \,,
\end{equation}
which casts with Eq. (17) after factorization into:
\begin{equation}
\left( \frac{\partial \bar{x}}{\partial \, T_{\rm rad}} \right)_{\rho} =
\frac{1}{T_{\rm rad}}\,\,\frac{\bar{x}\,\sum_{i} \nu_{i} x_{i} (1-x_{i}) \left( 1+ \frac{I_{i}}{k\, T_{\rm rad}} \right)}
{\sum_{i} \nu_{i} x_{i} (1-x_{i}) + \bar{x} } \,.
\end{equation}
We obtain by inserting Eq. (26) in Eq. (25) for every element $i$:
\begin{equation}
\left( \frac{\partial x_{i}}{\partial \, T_{\rm rad}} \right)_{\rho} =
\frac{1}{T_{\rm rad}}\,\, x_{i} (1-x_{i})\,\left( \left(1+\frac{I_{i}}{k\,T_{\rm rad}} \right)
- \frac{\sum_{i} \nu_{i} x_{i} (1-x_{i}) \left( 1+ \frac{I_{i}}{k\, T_{\rm rad}} \right)}
{\sum_{i} \nu_{i} x_{i} (1-x_{i}) + \sum_{i} \nu_{i} x_{i} }
\right)  \,.
\end{equation}
Equation (13) reduces, after inserting Eq. (26) and Eq. (27) and grouping terms, to:
\begin{eqnarray}
\left( \frac{{\partial} \, e}{{\partial} \, T_{\rm rad}} \right)_{\rho} = & & 
N\,k\left(  \left( \frac{3}{2}\,\theta\,\bar{x} - \sum_{i} \nu_{i} x_{i} (1-x_{i}) \frac{I_{i}}{k\, T_{\rm rad}} \right) 
 \frac{ \sum_{i} \nu_{i} x_{i} (1-x_{i}) \left(1+\frac{I_{i}}{k\,T_{\rm rad}}\right)}{ \bar{x} + \sum_{i}\nu_{i} x_{i} (1-x_{i})} \right)  \nonumber \\ 
+ & & N\, k\left( \sum_{i} \nu_{i} x_{i} (1-x_{i}) \frac{I_{i}}{k\, T_{\rm rad}} \left( 1+ \frac{I_{i}}{k\, T_{\rm rad}} \right) 
+ 12\,\alpha\,\theta\, (1+\bar{x})   \right) \,.
\end{eqnarray}
Hence, with Eq. (7) we obtain the total heat capacity at constant volume 
by summing Eq. (23) and Eq. (28), and grouping terms:
\begin{eqnarray}
\frac{c_{v}}{N\,k} & & = \left(\frac{3}{2} + 12\,\alpha\,\theta\right) (1+\bar{x}) 
+ \sum_{i} \nu_{i} x_{i} (1-x_{i}) \frac{I_{i}}{k\, T_{\rm e}}
\left( \frac{1}{2} + \theta \left(1+ \frac{I_{i}}{k\,T_{\rm rad}}\right) \right) \nonumber \\
& & +  \left( \frac{3}{2} \bar{x} -  
\sum_{i} \nu_{i} x_{i} (1-x_{i}) \frac{I_{i}}{k\,T_{\rm e}} \right) \frac{ \sum_{i}\nu_{i} x_{i} (1-x_{i}) 
\left( \frac{1}{2} + \theta \left(1+ \frac{I_{i}}{k\,T_{\rm rad}}\right) \right)   }{\bar{x} + 
\sum_{i} \nu_{i} x_{i} (1-x_{i})  } \,,
\end{eqnarray}   
where we normalize the specific heat capacity to the gas constant to obtain dimensionless units.
When denoting:
\begin{eqnarray} 
X=& & \sum_{i} \nu_{i} x_{i} (1-x_{i}) \,, \\
Y = & & \sum_{i} \nu_{i} x_{i} (1-x_{i}) \frac{I_{i}}{k\, T_{\rm e}} \,,
\end{eqnarray}
and we define the functions $Q$ and $G$ for element $i$:  
\begin{eqnarray}
Q_{i} = & & \frac{1}{2} +  \theta \left(1+ \frac{I_{i}}{k\, T_{\rm rad}} \right) \,, \\
G_{i} = & & Q_{i} \left( \frac{\frac{3}{2}\bar{x}-Y}{\bar{x}+X} + \frac{I_{i}}{k\,T_{\rm e}}  \right) \,,
\end{eqnarray}
Eq. (29) can formally be expressed:
\begin{equation}
\frac{c_{v}}{N\,k} = \left( \frac{3}{2} + 
12\,\alpha\,\theta \right)(1+\bar{x}) + \sum_{i} \nu_{i} x_{i} (1-x_{i}) \,G_{i} \,. 
\end{equation}
Equations (30)$-$(33) enable a similar compact form for the detailed expression 
of the heat capacity at constant total pressure, which we obtain below.

$iii$.  The kinetic temperature derivative of the ionization equation (5) at constant total pressure and radiation 
temperature yields after grouping terms and some rearrangement:
\begin{equation}
 \left( \frac{ \partial x_{i} }{\partial\, {\rm ln}\, T_{\rm e} }  \right)_{P_{\rm t}} =
x_{i}\,(1-x_{i})\,\left( \frac{3}{2} - \frac{1}{\bar{x}(1+\bar{x})}  \left( \frac{ \partial\, \bar{x} }{\partial \, {\rm ln} \, T_{\rm e} }  \right)_{P_{\rm t}} \right) \, 
 \,,
\end{equation}
which casts with Eq. (19) after factorization into:
\begin{equation}
 \left( \frac{ \partial\, \bar{x} }{\partial \, T_{\rm e} }  \right)_{P_{\rm t}} =
\frac{3}{2\,T_{\rm e}}\,\,\frac{\bar{x}(1+\bar{x})\,\sum_{i} \nu_{i} x_{i} (1-x_{i})}{\bar{x}(1+\bar{x}) + \sum_{i} \nu_{i} x_{i} (1-x_{i})} \,.
\end{equation}
We obtain by inserting Eq. (36) in Eq. (35) for every element $i$:
\begin{equation}
 \left( \frac{ \partial\, x_{i} }{\partial \, T_{\rm e} }  \right)_{P_{\rm t}} =
\frac{3}{2\,T_{\rm e}}\,\,\frac{x_{i}(1-x_{i}) \sum_{i} \nu_{i} x_{i} \left(1+\sum_{i} \nu_{i} x_{i}\right)}{ \sum_{i} \nu_{i} x_{i} (1-x_{i}) 
+\sum_{i} \nu_{i} x_{i} \left(1+\sum_{i} \nu_{i} x_{i} \right) } \,,
\end{equation}
for which we note the formal resemblance with Eq. (22), apart from an extra factor (\( 1+\sum_{i} \nu_{i} x_{i} \)) in the numerator 
and the second term of the denominator.  

$iv$. The kinetic temperature derivative of the density at constant total pressure and radiation temperature
results from differentiating the equation of state Eq. (3):
\begin{equation}
 \left( \frac{ \partial\, {\rm ln} \, \rho }{\partial \, {\rm ln} \, T_{\rm e} }  \right)_{P_{\rm t}} =
-\left(1+ \frac{1}{1+\bar{x}} \left(  \frac{ \partial\, \bar{x} }{\partial \, {\rm ln} \, T_{\rm e} } \right)_{P_{\rm t}}    \right) \,.
\end{equation}

$v$.  The radiation temperature derivative of the ionization equation (5) at constant total pressure and kinetic 
temperature yields after grouping terms and some rearrangement:
\begin{equation}
 \left( \frac{ \partial x_{i} }{\partial\, {\rm ln}\, T_{\rm rad} }  \right)_{P_{\rm t}} =
x_{i}\,(1-x_{i})\,\left( 1 + 4\alpha + \frac{I_{i}}{k\,T_{\rm rad}} - \frac{1}{\bar{x}(1+\bar{x})}  \left( \frac{ \partial\, \bar{x} }
{\partial \, {\rm ln} \, T_{\rm rad} }  \right)_{P_{\rm t}} \right) \, 
 \,,
\end{equation}
which casts with Eq. (19) after factorization into:
\begin{equation}
 \left( \frac{ \partial\, \bar{x} }{\partial \, T_{\rm rad} }  \right)_{P_{\rm t}} =
\frac{1}{T_{\rm rad}}\,\,\frac{\bar{x}(1+\bar{x})\,\sum_{i} \nu_{i} x_{i} 
(1-x_{i})\left(1+4\alpha+\frac{I_{i}}{k\,T_{\rm rad}} \right)}{\bar{x}(1+\bar{x}) + \sum_{i} \nu_{i} x_{i} (1-x_{i})} \,.
\end{equation}
We obtain by inserting Eq. (40) in Eq. (29) for every element $i$:
\begin{equation}
\left( \frac{\partial x_{i}}{\partial \, {\rm ln} \, T_{\rm rad}} \right)_{P_{\rm t}} =
\, x_{i} (1-x_{i})\,\left( 1+4\alpha+\frac{I_{i}}{k\,T_{\rm rad}} 
- \frac{\sum_{i} \nu_{i} x_{i} (1-x_{i}) \left( 1+ 4\alpha+ \frac{I_{i}}{k\, T_{\rm rad}} \right)}
{\sum_{i} \nu_{i} x_{i} (1-x_{i}) + \sum_{i} \nu_{i} x_{i}\left(1+\sum_{i} \nu_{i} x_{i}\right) }
\right)  \,,
\end{equation}
for which we note the formal resemblance with Eq. (27), apart from the extra terms with \( 4\alpha \) in the numerator,
and an extra factor (\(1+\sum_{i} \nu_{i} x_{i}\)) in the second term of the denominator.

$vi$.  The radiation temperature derivative of the density at constant total pressure and kinetic temperature
results from differentiating the equation of state Eq. (3):
\begin{equation}
 \left( \frac{ \partial\, {\rm ln} \, \rho }{\partial \, {\rm ln} \, T_{\rm rad} }  \right)_{P_{\rm t}} =
-\left(4\alpha + \frac{1}{1+\bar{x}} \left(  \frac{ \partial\, \bar{x} }{\partial \, {\rm ln} \, T_{\rm rad} } \right)_{P_{\rm t}}    \right) \,.
\end{equation}

Hence, with Eq. (10) we obtain the total heat capacity at constant total pressure 
by summing Eq. (14) and Eq. (15), and grouping terms:
\begin{eqnarray}
\frac{c_{P_{\rm t}}}{Nk} = & & \left( \frac{5}{2} + 16\,\alpha\, \theta - 4\,\alpha\, \left( \left( \frac{\partial\, {\rm ln} \, \rho}
{ \partial\, {\rm ln} \, T_{\rm e}}  \right)_{P_{\rm t}} + \theta \, \left( \frac{\partial\, {\rm ln} \, \rho}
{ \partial\, {\rm ln} \, T_{\rm rad}}  \right)_{P_{\rm t}}        \right) \right) \,  (1+\bar{x}) \nonumber \\
+ & &  \frac{5}{2} T_{\rm e} \left(  \left( \frac{\partial\, \bar{x}}
{ \partial\, T_{\rm e}}  \right)_{P_{\rm t}} + \left( \frac{\partial\, \bar{x}}
{ \partial\, T_{\rm rad}}  \right)_{P_{\rm t}}         \right) \nonumber \\
+ & &  \sum_{i} \nu_{i} \frac{I_{i}}{k} \left(  \left( \frac{\partial\, x_{i}}
{ \partial\, T_{\rm e}}  \right)_{P_{\rm t}} + \left( \frac{\partial\, x_{i}}
{ \partial\, T_{\rm rad}}  \right)_{P_{\rm t}}      \right) \,,
\end{eqnarray}
where we normalize the specific heat capacity to the gas constant to obtain dimensionless units.

When inserting Eqns. (36)$-$(38) and Eqns. (40)$-$(42) in Eq. (43), 
and defining with Eqns. (30)$-$(32) the function $H_{i}$ for element $i$:
\begin{equation}
H_{i}= \left( Q_{i} + 1 + 4\,\alpha\,\theta \right) \left( \frac{ (\frac{5}{2} + 4\,\alpha)\, \bar{x}\, (1+\bar{x}) - Y}
{ \bar{x}\, (1+\bar{x}) + X} + \frac{I_{i}}{k\,T_{\rm e}}\right) \,,
\end{equation}
we obtain, after grouping terms, the formal expression:
\begin{equation}
\frac{c_{P_{\rm t}}}{Nk} = \left( \frac{5}{2} + 4\,\alpha\,\left( 4\,\theta\,(\alpha+1)+1 \right) \right)  (1+\bar{x})
+ \sum_{i} \nu_{i} x_{i} (1-x_{i})\, H_{i} \,.
\end{equation}
\end{document}